\documentclass[twocolumn,longbibliography,aps,prb,superscriptaddress,showpacs]{revtex4-1}
\usepackage[latin9]{inputenc}
\usepackage{amsmath}
\usepackage{amssymb}
\usepackage[pdftex]{graphicx}
\usepackage{esint}

\begin{document}

\title{One-step approach to ARPES from strongly correlated solids: a Mott-Hubbard
system.}

\author{R.O.\ Kuzian}

\affiliation{Institute for Problems of Materials Science NASU, Krzhizhanovskogo
3, 03180 Kiev, Ukraine}
\affiliation{Donostia International Physics Center (DIPC), Paseo Manuel de Lardizabal 4, 
San Sebasti\'{a}n/Donostia, 20018 Basque Country, Spain}

\author{E.E.\ Krasovskii}
\affiliation{Departamento de F\'{i}sica de Materiales, 
Facultad de Ciencias Qu\'{i}imicas, Universidad del Pais Vasco/Euskal Herriko Unibertsitatea, 
Apdo. 1072, San Sebasti\'{a}n/Donostia, 20080 Basque Country, Spain}
\affiliation{Donostia International Physics Center (DIPC), Paseo Manuel de Lardizabal 4, 
San Sebasti\'{a}n/Donostia, 20018 Basque Country, Spain}
\affiliation{IKERBASQUE, Basque Foundation for Science, 48013 Bilbao, Spain}

\begin{abstract}

An expression is derived for angle-resolved photocurrent from a
semi-infinite correlated system. Within the sudden approximation, 
the photocurrent is proportional to the spectral function of a one-particle 
two-time retarded Green's function $\mathcal{G}$ of an operator that 
creates an electron in a special quantum state $\chi$ localized at the 
surface. For a system described by a many-body single-band model we present 
an analytical expression that relates the Green's function $\mathcal{G}$
with the Green's function of an infinite crystal 
$G_{b,\mathbf{k}}(\omega)$ in Wannier representation. 
The role of final states and of the crystal surface is analysed for a model 
Green's function of the infinite crystal with a three-peak spectral function 
typical of a Mott-Hubbard metal. The momentum dependence of both the 
quasiparticle pole position and the spectral weight of the incoherent 
band manifest themselves in the shape of the photocurrent energy 
distribution curve.

\end{abstract}

\date{\today}


\maketitle

\section{Introduction}

Angle-resolved photoemission spectroscopy (ARPES) has proved an indispensable 
tool to study the electronic structure of solids 
\cite{Hufner03,Damascelli03,Schattke08b,Kordyuk14}.
It has become especially important with the discovery of high-Tc cuprate 
superconductors, when the enhanced experimental and theoretical effort
was put into the studies of strongly correlated electron systems (SCES)
\cite{Fulde91,Imada98}. Mean-field-based approaches fail to describe
the valence band of SCES, so ARPES becomes a crucial source
of information about the electronic structure and a verification tool
for many-body theories \cite{Sawatzky89}. However, the interpretation
of ARPES in terms of one-electron many-body spectral function (SF)
may be sufficient only when the energy dispersion perpendicular to
the surface is of the order of or smaller than the experimental energy
resolution. This is the case in the layered cuprates \cite{Damascelli03,Kordyuk14}
or perovskite-type vanadates \cite{Takizawa09,Yoshida10,Aizaki12,Laverock15},
which have a quasi-two-dimensional valence and conduction bands despite
their cubic lattice. Still, most of correlated compounds have a three-dimensional
electronic structure \cite{Imada98}, and the photohole dispersion
normal to the surface requires a more thorough theoretical analysis
of the ARPES.

A conclusive interpretation of the ARPES experiment depends on the
knowledge of final states of the photoemission process. In the sudden
approximation \cite{Borstel85}, the final states are time-reversed
low-energy electron diffraction (LEED) states \cite{Adawi64}, which
decay into the interior of the solid in accord with the surface sensitivity
of ARPES. To be realistic, a proper calculation of the photocurrent should allow
for changes of the electronic structure near the surface
and include the excitation probabilities. The most elaborate
theoretical framework to deal with ARPES is the one-step theory
\cite{Adawi64,Mahan70,Caroli73,Feibelman74,Pendry76,Jepsen82}.
It describes the excitation, the transport of the photoelectron to
the crystal surface, and the escape into the vacuum as a single quantum-mechanical
process including all multiple-scattering events. This approach allows
to perform realistic photocurrent calculations based on Kohn-Sham
eigenfunctions, and it is implemented in several computer codes 
\cite{Pendry76,Braun96,Krasovskii04}.

The one-step approach was also formulated for nonlocal potentials
\cite{Potthoff97,Meyer99}, and in Refs. \onlinecite{Minar05,Minar11}
it was combined with the dynamic mean field theory (DMFT) \cite{Georges96,Kotliar06,Vollhardt12}
within the Korringa-Kohn-Rostoker multiple scattering formalism. It
was recently applied for the interpretation of photoemission spectra
of 3$d$ metals \cite{Braun06,SanchesBarriga09,SanchesBarriga12}.
Most of studies of SCES are, however, performed within the basis of
localized Wannier functions\cite{Wannier37,Marzari97,Marzari12},
as originally proposed by P.W. Anderson \cite{Anderson59} and J.
Hubbard \cite{Hubbard63}. The Wannier representation is quite natural
here, since the largest Coulomb interaction term -- the so-called
Hubbard interaction, which is responsible for electron correlations
in $d$ or $f$ shells of transition metals -- is diagonal in this
basis \cite{Hubbard63}. The theoretical many-body bulk SFs are often
directly compared with ARPES spectra \cite{Nekrasov06,Byczuk07,Kuchinskii12},
thus ignoring the role of the final states and the effect of the
surface. The surface effects in SCES were considered in 
Refs.~\onlinecite{Potthoff96,Potthoff99,Liebsch03,Ishida09,Nourafkan11},
and their influence on ARPES was discussed on a qualitative level.
Thus, the formulation of the one-step approach in the localized basis
is highly desirable as it would enable a quantitative comparison of
many-body calculations results with the state-of-the-art ARPES data
\cite{Borisenko12}.

According to the classification of Ref.~\onlinecite{Zaanen85},
the strongly correlated transition metal compounds may be divided
into two categories depending on the relation between Coulomb interaction
$U$ within the $d$-shell and the charge transfer energy $\Delta$
between the metal ion and surrounding anions. In the Mott-Hubbard
systems, $\Delta\gg U$, the valence band may be described by a one-band
Hubbard-type model \cite{Hubbard63}, while for charge-transfer systems,
$\Delta\ll U$, an explicit account of the anion states is necessary
\cite{Emery87}.

In this paper, we formulate the one-step approach in the localized
basis (Sec. \ref{sec:ARPES-local}), and consider its application
to Mott-Hubbard systems. In Sec. \ref{sec:surf}, starting from the
bulk Green's function (GF), we derive the GF of the semi-infinite
system. After a short discussion of ARPES of layered systems in Sec.
\ref{sec:2D}, we find an analytical formula for the photocurrent
from a system with tangible dispersion in the direction normal to
the surface (Sec. \ref{sec:3D}). In Sec. \ref{sec:Discussion}, we
discuss how the formula reflects the role of final states and of the
surface and give some illustration of its application. Technical details
of the derivation are given in the Appendix.

\section{ARPES calculation for localized basis\label{sec:ARPES-local}}

In this section we revisit the one-step theory of photoemission in
order to formulate it in the Wannier representation for initial states.
The localized basis is ideally suited for the electronic structure
of SCES, and at a certain level of approximation it allows an elegant
inclusion of surface effects.

\subsection{Sudden approximation}

We consider a semi-infinite crystal that extends over the half-space
$z\leq z_{0}$, with a perfectly flat surface. The solid is irradiated
with light given by the vector potential 
$\mathbf{A}\left(\mathbf{x},t\right)=\mathbf{A}\left(\mathbf{x}\right)\cos\Omega t$
(we choose the gauge in which the scalar potential is zero). Within
the sudden approximation \cite{Borstel85}, the interaction between
the excited electron and the photohole is neglected. Then the descriptions
of the initial state and of the final state can be separated from
each other. The steady radial photocurrent $j\left(\hat{q},E\right)$
of electrons emerging from the solid along the observation direction
defined by the unit vector $\hat{q}$ with energies between
$E$ and $E+dE$ is then given by\cite{Caroli73,Feibelman74} 
\begin{align}
j\left(\hat{q},E\right) & =\frac{1}{2\pi}\lim_{\substack{\mathbf{X}^{\prime}
\to \mathbf{X},\\ \mathbf{X}\rightarrow\infty}}
\left(\frac{\partial}{\partial X^{\prime}}-\frac{\partial}{\partial X}\right)\nonumber \\
 & \times\iint d^{3}\mathbf{x}_{1}d^{3}\mathbf{x}_{2}
 G\left(\mathbf{X},\mathbf{x}_{1},E\right)\hat{O}\left(\mathbf{x}_{1}\right)\nonumber \\
 & \times G^{+}\left(\mathbf{x}_{1},\mathbf{x}_{2},E
 -\hbar\Omega\right)\hat{O}\left(\mathbf{x}_{2}\right)G^{*}
 \left(\mathbf{x}_{2},\mathbf{X}^{\prime},E\right),\label{eq:j}
\end{align}
where the vector $\mathbf{X}=X\hat{q}$ points in the direction of
the detector, and $G\left(\mathbf{X},\mathbf{x},E\right)$ is the
retarded propagator of the outgoing electron, 
\begin{equation*}
G(\mathbf{x},\mathbf{x}^{\prime},\omega) =\left\langle \!\left\langle 
\hat{\psi}(\mathbf{x})|\hat{\psi}^{\dagger}(\mathbf{x}^{\prime})\right\rangle 
\!\right\rangle _{\omega}.
%\label{eq:Grrp}
\end{equation*}
In general, the anticommutator two-time retarded GF of two operators
$\hat{A}$ and $\hat{B}$ is defined as 
\begin{equation*}
\left\langle \!\left\langle \hat{A}|\hat{B}\right\rangle 
\!\right\rangle _{\omega}\equiv-i\int_{0}^{\infty}\left\langle 
\left\{ \hat{A}(t),\hat{B}(0)\right\} \right\rangle e^{i\omega t}dt,
%\label{eq:ttGF}
\end{equation*}
where $\left\{ \hat{A},\hat{B}\right\} \equiv\hat{A}\hat{B}+\hat{B}\hat{A}$,
the time-dependent operator $\hat{A}(t)$ is 
$\hat{A}(t)=\exp(it\hat{H})\hat{A}\exp(-it\hat{H})$,
and the angular brackets denote the thermodynamic average 
$\left\langle \hat{A}\right\rangle \equiv Tr\left[\exp\left(-\beta\hat{H}
\right)\hat{A}\right]/Tr\left[\exp\left(-\beta\hat{H}\right)\right]$.
The operator $\psi(\mathbf{x})$ annihilates an electron at the point
$\mathbf{x}$, and $G^{+}\left(\mathbf{x}_{1},\mathbf{x}_{2},\omega\right)$
is the ``lesser'' function \cite{Caroli73,Feibelman74} for the
initial state 
\begin{equation}
G^{+}\left(\mathbf{x}_{1},\mathbf{x}_{2},\omega\right) \equiv
-2if\left(E+\Phi\right)G^{\prime\prime}\left(\mathbf{x}_{1},
\mathbf{x}_{2},\omega\right),
\label{eq:Gpls}
\end{equation}
where $\Phi$ is the work function, the vacuum level is at $E=0$,
and $f(\omega)=1/\left(e^{\beta\omega}+1\right)$ is the Fermi distribution
function. We consider a non-magnetic solid and drop the spin index. 
Throughout the text the double prime denotes the imaginary part of a 
complex value, e.g., $G^{\prime\prime}\equiv\mathrm{Im}\,G$. 
The operator 
$
\hat{O}\left(\mathbf{x}\right)=\frac{1}{2\mathsf{c}}\left[\mathbf{A}
\left(\mathbf{x}\right)\cdot\mathbf{P}+\mathbf{P}\cdot\mathbf{A}
\left(\mathbf{x}\right)\right]
$
is the electron-light coupling, with $\mathbf{P}=-i\nabla$ being
the electron momentum operator and $\mathsf{c}$ the light velocity.
The atomic units $\hbar=e=m_{e}=1$ are used. In resonant photoemission
\cite{Tanaka94,Kay01,DaPieve13,DaPieve16} or in the presence of microscopic
fields due to the dielectric screening \cite{Feibelman1975,Krasovskii2010}
the operator $\hat{O}$ is more involved, which complicates the calculation
of matrix elements $\mathsf{M}\left(\mathbf{k}_{\parallel},E\right)$
in Eq.~(\ref{eq:M}), but the theory presented below remains fully
applicable

Following Ref. \onlinecite{Feibelman74}, we use the asymptotic
formula for $G\left(\mathbf{X},\mathbf{x},E\right)$, 
\begin{equation}
G\left(\mathbf{X},\mathbf{x},E\right)\xrightarrow[\mathbf{X}\to \infty]{}
\frac{1}{2\pi}\frac{\exp\left(iX\sqrt{2E}\right)}{X}\varphi_{>}
\left(\mathbf{x},\hat{q},E\right),\label{eq:Grass}
\end{equation}
where $\varphi_{>}\left(\mathbf{x},\hat{q},E\right)$ is the LEED
wave function. The inelastic scattering due to electron-electron interaction
in the propagation of the outgoing electron may be taken into account
phenomenologically by introducing an absorbing optical potential into
the effective Schr\"odinger equation for the function 
$\varphi_{>}\left(\mathbf{x},\hat{q},E\right)$
\cite{Slater37,Strocov97,Barrett05}. Thereby the LEED function becomes
a superposition of evanescent Bloch waves. Substitution of (\ref{eq:Grass})
and (\ref{eq:Gpls}) into Eq.~(\ref{eq:j}) gives \begin{widetext}
\begin{align}
j\left(\hat{q},E\right) & =-\left(\frac{1}{2\pi}\right)^{3}
\left[\frac{2f\left(E+\Phi\right)\sqrt{2E}}{X^{2}}\right]\nonumber \\
 & \times\iint d^{3}\mathbf{x}_{1}d^{3}\mathbf{x}_{2}\varphi_{>}
 \left(\mathbf{x}_{1},\hat{q},E\right)\hat{O}\left(\mathbf{x}_{1}\right)
 G^{\prime\prime}\left(\mathbf{x}_{1},\mathbf{x}_{2},E-\hbar\Omega\right)
 \hat{O}\left(\mathbf{x}_{2}\right)\varphi_{>}^{*}\left(\mathbf{x}_{2},
 \hat{q},E\right).\label{eq:j2}
\end{align}
\end{widetext}

Note that the initial states are confined inside the solid, so that
the integration over $\mathbf{x}_{1}$ and $\mathbf{x}_{2}$ in (\ref{eq:j2})
is essentially restricted to the crystal half-space, $\mathbf{x}_{i}\subset\mathcal{S}$,
i.e., 
\[
\int_{\mathbf{x}\subset\mathcal{S}}d^{3}\mathbf{x}_{1}\ldots\equiv
\iint_{-\infty}^{\infty}dxdy\int_{-\infty}^{z_{0}+\Delta z}dz\ldots,
\]
which assumes that initial states vanish at a distance $\Delta z$
from the surface. With this in mind, and using the symmetry relation
\begin{equation}
G(\mathbf{x}_{1},\mathbf{x}_{2},\omega)=G(\mathbf{x}_{2},\mathbf{x}_{1},\omega),
\label{eq:Geq}
\end{equation}
we make an important next step (the details are given in the Appendix),
and rewrite (\ref{eq:j2}) in the form 
\begin{align}
j\left(\hat{q},E\right) & =\left[\frac{f\left(E+\Phi\right)\sqrt{2E}}{
\left(2\pi X\right)^{2}}\right]\mathcal{A}\left(\hat{q},E-\hbar\Omega\right),
\label{eq:jCCdag}\\
\mathcal{A}\left(\hat{q},\omega\right) & =-\frac{1}{\pi}\mathrm{Im}\mathcal{G}
\left(\hat{q},\omega+i0\right),\label{eq:ACCdag}\\
\mathcal{G}\left(\hat{q},\omega\right) & =\left\langle \!\left\langle 
\hat{C}|\hat{C}^{\dagger}\right\rangle \!\right\rangle _{\omega},
\label{eq:GCCdag}
\end{align}
where the operator
\begin{equation}
\hat{C}^{\dagger}\left(\hat{q},E\right)\equiv\int d^{3}\mathbf{x}
\hat{\psi}^{\dagger}(\mathbf{x})\chi\left(\mathbf{x},\hat{q},E\right)
\label{eq:C}
\end{equation}
creates an electron in a state with the wave function 
\begin{align}
\chi\left(\mathbf{x},\hat{q},E\right) & =\hat{O}\left(\mathbf{x}\right)
\varphi_{>}^{\ast}\left(\mathbf{x},\hat{q},E\right),
\:\mathbf{x}\subset\mathcal{S}\label{eq:chiC}\\
 & =0,\:\textrm{otherwise}\nonumber 
\end{align}
Equation (\ref{eq:ACCdag}) gives an explicit form of the
SF to be calculated to obtain the photocurrent.

\subsection{Non-interacting electrons}

In a mean-field approach, the initial states are described by an effective
one-particle Hamiltonian. In the basis of its eigenfunctions 
$\Psi_{i}\left(\mathbf{x}\right)$,
it reads 
\begin{equation}
\hat{H}_{\mathrm{mf}}=\sum_{i}E_{i}a_{i}^{\dagger}a_{i},\label{eq:Hmf}
\end{equation}
where index $i$ incorporates all quantum numbers that define a quantum
state of the system, $E_{i}$ being its energy. Then the electron
annihilation operator is 
$
\hat{\psi}\left(\mathbf{r}\right)=\sum_{i}\Psi_{i}\left(\mathbf{x}\right)
a_{i},
$
and the operator conjugate to $\hat{C}^{\dagger}$ of Eq.~(\ref{eq:C}) is
\begin{align*}
\hat{C}\left(\hat{q},E\right) & =\sum_{i}M_{i}a_{i},\nonumber \\
M_{i} & =\int_{\mathbf{x}\subset\mathcal{S}}d^{3}\mathbf{x}
\varphi_{>}\left(\mathbf{x},\hat{q},E\right)\hat{O}\left(\mathbf{x}\right)
\Psi_{i}\left(\mathbf{x}\right)\nonumber \\
 & =\int d^{3}\mathbf{x}\chi^{*}\left(\mathbf{x},\hat{q},E\right)
 \Psi_{i}\left(\mathbf{x}\right).
\end{align*}
The last integration may be extended over the whole 
space, as both functions $\chi$ and $\Psi$ are confined inside the solid. The 
GF of Eq.~(\ref{eq:GCCdag}) that defines the photocurrent is 
\[
\mathcal{G}\left(\hat{q},\omega\right)=\sum_{i}M_{i}M_{j}^{*}
\left\langle \!\left\langle a_{i}|a_{j}^{\dagger}\right\rangle \!\right\rangle _{\omega}.
\]
The GFs in the right-hand side are trivially calculated 
\begin{equation}
G_{ij}(\omega)\equiv\left\langle \!\left\langle a_{i}|a_{j}^{\dagger}
\right\rangle \!\right\rangle _{\omega}=\frac{\delta_{ij}}{\omega-E_{i}}.\label{eq:GFaamf}
\end{equation}
We see that the SF (\ref{eq:ACCdag}) reduces to the density of states
(DOS) projected on the function $\chi\left(\mathbf{x},\hat{q},E\right)$
\begin{equation}
\mathcal{A}\left(\hat{q},\omega\right)=\sum_{i}\left|M_{i}\right|^{2}
\delta\left(\omega-E_{i}\right).\label{eq:Amf}
\end{equation}
Substituting Eq.~(\ref{eq:Amf}) into Eq.~(\ref{eq:jCCdag}), 
we recover the well-known expression for the photocurrent in the mean-field 
one-step approach \cite{Adawi64,Mahan70,Caroli73,Feibelman74,Pendry76,
Jepsen82,Braun96,Krasovskii97,Cui10}.
Note that the Hamiltonian (\ref{eq:Hmf}) describes a \emph{semi-infinite}
crystal, which makes the calculation of the eigenvalues $E_{i}$ and
eigenfunctions $\Psi_{i}\left(\mathbf{x}\right)$ highly non-trivial
even in the mean-field approximation.

\subsection{Interacting electrons}

For the non-interacting systems, the photoexcitation of an electron
from a single-determinant $N$-electron eigenstate of the Hamiltonian
(\ref{eq:Hmf}) creates an $(N-1)$-electron eigenstate of the same
Hamiltonian. The electron-electron interaction complicates the picture
of the photoexcitation. On the mean-field level, the removal of an
electron from an $N$-electron system changes the mean-field, but
these changes are negligible for a macroscopic number of electrons.
More important is the interaction beyond the mean-field: the two-particle
(four-fermion operator) terms in the Hamiltonian, which account for
the residual interaction, i.e., are the part of the bare Coulomb interaction
responsible for the correlations in the electron motion \cite{Fulde91}.
In contrast to the bare Coulomb interaction, the residual interaction
is a short-ranged one. It is often introduced on the model level via
Hubbard-like terms, which are conveniently represented in the localized
basis of Wannier functions \cite{Hubbard63}.

In SCES, the Hubbard terms are comparable with matrix elements of
kinetic energy. This makes it impossible to present an $N$-electron
eigenstate as a single determinant. The removal/addition of an electron
from/to this state produces an $(N-1)$-/($N+1$)-electron state that
is a combination of a large number of eigenstates with different energies.
As a consequence, the GF describing electron removal and additional
spectra does not have the simple pole form of Eq.~(\ref{eq:GFaamf})
but acquires a complex self-energy in the denominator. As a result,
the SF $-G_{ii}^{\prime\prime}(\omega+i0)/\pi$ is not anymore a single
$\delta$-function, but it may have humps that come from the branch
cut singularities of the self-energy, and that are called incoherent
bands. These bands coming from the self-energy of the \emph{initial}
states are observed in ARPES as ``satellites'' that appear at binding
energies different from the energies of ``main peaks'' of the mean
field theory. For a proper interpretation of the experiment, the many-body
GF describing the initial states should be incorporated into the one-step
approach.

Note that the only approximation we have used to derive 
Eq.~(\ref{eq:jCCdag}) is the sudden approximation, and that Eq.~(\ref{eq:jCCdag})
is fully general and applicable for a wide range of systems including
strongly correlated systems. The role of the final state 
$\varphi_{>}^{\ast}\left(\mathbf{x},\hat{q},E\right)$
(the time reversed LEED state) is clear from Eqs. (\ref{eq:C}) and
(\ref{eq:chiC}): it defines the form of the operator 
$\hat{C}_{\sigma}\left(\hat{q},E\right)$,
and, thus, the SF (\ref{eq:ACCdag}), which is our ultimate aim. Thus,
the angular and energy dependence of the photocurrent cannot 
be understood solely from the structure of the initial states. On the 
one hand, this complicates the interpretation of ARPES experiments, 
but, on the other hand, it allows to learn about final states from 
the measured spectra \cite{Krasovskii07}.

We assume that the target crystal has two-dimensional (2D) lattice
periodicity. Inside the solid, the time reversed LEED function may
be written as 
\begin{equation}
\varphi_{>}^{*}\left(\mathbf{x},\hat{q},E\right)=e^{i\mathbf{k}_{\parallel}
\mathbf{x}_{\parallel}}U\left(\mathbf{x}_{\parallel},z,\hat{q},E\right),
\label{eq:LEED}
\end{equation}
where $\mathbf{x}_{\parallel}$ is the radius-vector component parallel
to the surface, $\mathbf{x}=\mathbf{x}_{\parallel}+z\mathbf{n}$,
with $\mathbf{n}$ being the unity vector normal to the surface. The
surface-parallel momentum component 
$\mathbf{q}_{\parallel}=\mathbf{k}_{\parallel}+\mathbf{G}_{\parallel}$
is the sum of the momentum vector in the first Brillouin zone $\mathbf{k}_{\parallel}$
and 2D reciprocal lattice vector $\mathbf{G}_{\parallel}$. The function
$U\left(\mathbf{x}_{\parallel},z,\hat{q},E\right)$ is periodic in
$\mathbf{x}_{\parallel}$ and may be written as a combination of evanescent
waves {[}cf. Eq. (37) of Ref. \onlinecite{Feibelman74}{]}: 
\begin{align}
U\left(\mathbf{x}_{\parallel},z,\hat{q},E\right) & =
\sum_{m}\varphi_{m}^{*}\left(\mathbf{x},\mathbf{k}_{\parallel},E\right),
\label{eq:Ufim}\\
\varphi_{m}^{*}\left(\mathbf{x},\mathbf{k}_{\parallel},E\right) &
 \equiv e^{ik_{\perp,m}\left(z-z_{0}\right)}u_{m}\left(\mathbf{x},
 \mathbf{k}_{\parallel},E\right),\label{eq:fim}
\end{align}
where $m$ is the band index and 
$k_{\perp,m}\left(E,\mathbf{k}_{\parallel}\right)=
k_{z,m}^{\prime}-ik_{z,m}^{\prime\prime}$
($k_{z,m}^{\prime},k_{z,m}^{\prime\prime}>0$) is the complex momentum
component in the direction perpendicular to the surface.
The functions $u_{m}$ have the periodicity of the 3D crystal, with 
the Bravais lattice vectors $\mathbf{R}=\mathbf{R}_{\parallel}+lc\mathbf{n}$,
where $l\leq z_{0}/c$ is an integer and $c$ is the lattice period in $z$ 
direction. For both $\mathbf{x}$ and $\mathbf{x}+\mathbf{R}$ inside 
the solid it is 
$u_{m}\left(\mathbf{x}+\mathbf{R},\mathbf{k}_{\parallel},E\right)
=u_{m}\left(\mathbf{x},\mathbf{k}_{\parallel},E\right)$.
Inside the crystal, the function produced
by the perturbation $\hat{O}$ acting on the final state $\varphi_{>}^{*}$
in Eq.~(\ref{eq:chiC}) is, clearly, also a combination of evanescent
waves 
\begin{align*}
\chi\left(\mathbf{x},\hat{q},E\right) & =
\sum_{m}\chi_{m}\left(\mathbf{x},\mathbf{k}_{\parallel},E\right),\\
\chi_{m}\left(\mathbf{x},\mathbf{k}_{\parallel},E\right) & 
=\hat{O}\left(\mathbf{x}\right)e^{i\mathbf{k}_{\parallel}\mathbf{x}_{\parallel}}
\varphi_{m}^{*}\left(\mathbf{x},\mathbf{k}_{\parallel},E\right).
\end{align*}
Thus, the function $\chi\left(\mathbf{x},\hat{q},E\right)$ is localized 
at the surface, which reflects the surface sensitivity
of the photoemission spectroscopy. If one of the waves dominates the
sum in Eq. (\ref{eq:Ufim}) its localization can be expressed by the
``inelastic mean free path'' 
$L\sim1/2k_{z}^{\prime\prime}\left(E,\mathbf{k}_{\parallel}\right)$.

In order to proceed further, we chose a basis of localized functions
suitable for the description of the initial state. For example, it
may be the basis of Wannier functions for a set of bands within some
energy window around $\omega=E-\hbar\Omega$. We write the electron
annihilation operator for the initial state in the form 
\begin{equation*}
\hat{\psi}(\mathbf{r})=\sum_{\mathbf{R},\alpha}w_{\alpha}
\left(\mathbf{r}-\mathbf{R}-\mathbf{s}\right)a_{\mathbf{R},\alpha},
%\label{eq:psicrest}
\end{equation*}
where $a_{\mathbf{R},\alpha}$ annihilates an
electron in the state $w_{\alpha}(\mathbf{r}-\mathbf{R}-\mathbf{s})$
localized at the lattice site $\mathbf{R}+\mathbf{s}$, where $\mathbf{s}$
is a basis vector of the unit cell, and $\alpha$ accumulates $\mathbf{s}$ 
and all the relevant quantum numbers. For
the operator $\hat{C}$ of Eq.~(\ref{eq:C}) we obtain (see Appendix)
\begin{widetext} 
\begin{align}
\hat{C}\left(\hat{q},E\right) & =\sqrt{N_{\parallel}}\sum_{l,\mathbf{s},
\alpha}\int dz\iint d^{2}\mathbf{x}_{\parallel}\varphi_{>}\left(\mathbf{x},
\hat{q},E\right)\hat{O}\left(\mathbf{x}\right)w_{\alpha}
\left[\mathbf{x}_{\parallel}+\left(z-lc\right)\mathbf{n}-
\mathbf{s}\right]a_{\mathbf{k}_{\parallel},l,\alpha}\label{eq:Ckl}\\
 & =\sqrt{N_{\parallel}}\sum_{m,\alpha}\mathsf{M}_{m,\alpha}
 \left(\mathbf{k}_{\parallel},E\right)\sum_{l}e^{-ik_{\perp,m}^{*}
 \left(lc-z_{0}\right)}a_{\mathbf{k}_{\parallel},l,\alpha},\label{eq:CklM}\\
 & =\sqrt{\frac{N_{\parallel}}{N_{\perp}}}\sum_{m,\alpha}
 \mathsf{M}_{m,\alpha}\left(\mathbf{k}_{\parallel},E\right)\sum_{p}
 e^{ipz_{0}}\Delta_{m,p}a_{\mathbf{k}_{\parallel},p,\alpha},\label{eq:Ck}\\
\mathsf{M}_{m,\alpha}\left(\mathbf{k}_{\parallel},E\right) & \equiv\int
d^{3}\mathbf{x}\chi_{m}^{*}\left(\mathbf{x},\mathbf{k}_{\parallel},E\right)
w_{\alpha}\left[\mathbf{x}-z_{0}\mathbf{n}-\mathbf{s}\right],\label{eq:M}\\
\Delta_{m,p} & \equiv\sum_{l=-\infty}^{z_{0}/c}e^{-i\left(k_{\perp,m}^{*}
-p\right)\left(lc-z_{0}\right)}=\left\{ 1-e^{i\left(k_{\perp,m}^{*}
-p\right)c}\right\} ^{-1},\label{eq:Dlt}
\end{align}
\end{widetext}
where we have introduced the Fourier transforms
\begin{align}
a_{\mathbf{k}_{\parallel},l,\alpha} & =\frac{1}{\sqrt{N_{\parallel}}}
\sum_{\mathbf{R}_{\parallel}}\mathrm{e}^{-i\mathbf{k}_{\parallel}
\mathbf{R}_{\parallel}}a_{\mathbf{R}_{\parallel},l,\alpha},\label{eq:akapz}\\
a_{\mathbf{k},\alpha} & =a_{\mathbf{k}_{\parallel},p,\alpha}
=\frac{1}{\sqrt{N_{\perp}}}\sum_{l=-\infty}^{\infty}\mathrm{e}^{-iplc}
a_{\mathbf{k}_{\parallel},l,\alpha}.\label{eq:ak}
\end{align}
Here $N_{\parallel}$ is the number of sites in the plane, $N_{\perp}$
is the number of planes in the system, and 
$\mathbf{k}=\mathbf{k}_{\parallel}+p\mathbf{n}$.
Operator $a_{\mathbf{k}_{\parallel},l,\alpha}$ annihilates
an electron in a layer Bloch state 
\begin{align}
 & w_{\mathbf{k}_{\parallel},\alpha}\left(\mathbf{r}
 -lc\mathbf{n}\right)= \nonumber \\
 & \frac{1}{\sqrt{N_{\parallel}}}\sum_{\mathbf{R}_{\parallel}}
 \mathrm{e}^{i\mathbf{k}_{\parallel}\mathbf{R}_{\parallel}}w_{\alpha}
 \left[\mathbf{x}_{\parallel}-\mathbf{R}_{\parallel}+\left(z-lc\right)
 \mathbf{n}-\mathbf{s}\right],\label{eq:fikapl}
\end{align}
localized at $l$-th layer, while $a_{\mathbf{k},\mathbf{s},\alpha}$
annihilates an electron in a bulk Bloch state 
\begin{align*}
 & w_{\mathbf{k},\alpha}\left(\mathbf{r}\right)=
 \frac{1}{\sqrt{N_{\perp}}}\sum_{l}\mathrm{e}^{iplc}
 w_{\mathbf{k}_{\parallel},\alpha}\left(\mathbf{r}-lc\mathbf{n}\right)\nonumber \\
 & =\frac{1}{\sqrt{N_{\parallel}N_{\perp}}}\sum_{\mathbf{R}}\mathrm{e}^{i\mathbf{k}
 \mathbf{R}}w_{\alpha}\left(\mathbf{r}-\mathbf{R}-\mathbf{s}\right).
 %\label{eq:fik}
\end{align*}
Equation (\ref{eq:Ckl}) expresses the conservation of the momentum
parallel to the surface. The expression (\ref{eq:Ck}) for $\hat{C}$
shows that, generally, all states with different perpendicular momenta
$p$ contribute to the photocurrent for a given $\mathbf{k}_{\parallel}$
and $E$. In Ref.\ \onlinecite{Feibelman74}, it was pointed out
that the factor $\Delta_{m,p}$ (\ref{eq:Dlt}) is sharply peaked
at $k_{\perp,m}^{\prime}=p$ if $k_{\perp,m}^{\prime\prime}c\ll1$
{[}cf. Eqs. (42)-(47) of Ref.\ \onlinecite{Feibelman74}{]}. In
this particular case, the crystal momentum is conserved also in $z$
direction. In Sec.~\ref{sec:3D} we will return to this discussion. 

Equations (\ref{eq:CklM}) and (\ref{eq:Ck}) allow to write the GF of 
Eq.~(\ref{eq:GCCdag}) in the form
\begin{align}
 & \mathcal{G}\left(\hat{q},\omega\right)=
 \frac{N_{\parallel}}{N_{\perp}}\sum_{\substack{\substack{m}
_{1},m_{2},\backslash\alpha_{1},\alpha_{2}}
}\mathsf{M}_{m_{1},\alpha_{1}}\mathsf{M}_{m_{2},\alpha_{2}}^{*}\nonumber \\
 & \times\sum_{l_{1},l_{2}}e^{ik_{\perp,m_{2}}\left(l_{2}c-z_{0}\right)
 -ik_{\perp,m_{1}}^{*}\left(l_{1}c-z_{0}\right)}G_{\mathbf{k}_{\parallel},
 l_{1},l_{2},\alpha_{1},\alpha_{2}}(\omega)\label{eq:Gl1l2calG}\\
 & =\frac{N_{\parallel}}{N_{\perp}}\sum_{\substack{\substack{m}
_{1},m_{2},\backslash\alpha_{1},\alpha_{2}}
}\mathsf{M}_{m_{1},\alpha_{1}}\mathsf{M}_{m_{2},\alpha_{2}}^{*}\nonumber \\
 & \times\sum_{p_{1},p_{2}}e^{i\left(p_{1}-p_{2}\right)z_{0}}
 \Delta_{m_{1},p_{1}}\Delta_{m_{2},p_{2}}^{*}G_{\mathbf{k}_{\parallel},
 p_{1},p_{2},\alpha_{1},\alpha_{2}}\left(\omega\right),\label{eq:Gp1p2calG}\\
 & G_{\mathbf{k}_{\parallel},l_{1},l_{2},\alpha_{1},\alpha_{2}}(\omega)
 \equiv \left\langle \!\left\langle a_{\mathbf{k}_{\parallel},l_{1},
 \alpha_{1}}|a_{\mathbf{k}_{\parallel},l_{2},\alpha_{2}}^{\dagger}
 \right\rangle \!\right\rangle _{\omega}.\label{eq:Gl1l2a1a2}
\end{align}
The GF of semi-infinite crystal 
\begin{equation}
G_{\mathbf{k}_{\parallel},p_{1},p_{2},\alpha_{1},\alpha_{2}}\left(\omega\right)
\equiv\left\langle \!\left\langle a_{\mathbf{k}_{\parallel},p_{1},
\alpha_{1}}|a_{\mathbf{k}_{\parallel},p_{2},\alpha_{2}}^{\dagger}\right
\rangle \!\right\rangle _{\omega}\label{eq:Gp1p2a1a2}
\end{equation}
depends on the pair of perpendicular momenta because of the broken translational
invariance in the surface-normal direction.

\section{Semi-infinite Mott-Hubbard system\label{sec:surf}}

Now we consider a system whose valence band spectrum may be described
by an effective one-band Hamiltonian $\hat{H}_{\mathrm{eff}}$ on
a Bravais lattice, i.e., we have only one sort of orbitals 
$\phi(\mathbf{r}-\mathbf{R})$
 at the lattice sites $\mathbf{R}=\mathbf{R}_{\parallel}+z\mathbf{n}$.
For an infinite crystal, the GF is diagonal in the $\mathbf{k}$-space 
\begin{align}
\left\langle \!\left\langle a_{\mathbf{k}_{1}}|
a_{\mathbf{k_{2}}}^{\dagger}\right\rangle \!\right\rangle _{\omega} &
 =\delta_{\mathbf{k_{1},k}_{2}}G_{b,\mathbf{k}}(\omega),\label{eq:aadagg}\\
G_{b,\mathbf{k}}(\omega) & =\frac{1}{\omega-\varepsilon_{\mathbf{k}}
-\Sigma_{\mathbf{k},\omega}},\label{eq:Gkkb}
\end{align}
where $\mathbf{k}=\mathbf{k}_{\parallel}+p\mathbf{n}$. Here we do
not specify the Hamiltonian $\hat{H}_{\mathrm{eff}}$ but assume only
that the mean-field energy $\varepsilon_{\mathbf{k}}$ and the self-energy
$\Sigma_{\mathbf{k},\omega}$ may be calculated for the bulk system
with the full account of many-body effects. The momentum-dependent
SF 
\begin{equation}
A_{\mathrm{b}}(\mathbf{k},\omega+i0)=-\mathrm{Im}G_{b,\mathbf{k}}(\omega+i0)/\pi
\label{eq:Ab}
\end{equation}
is the main characteristic of the electronic structure of SCES. It
contains information both about the quasiparticle energy dispersion
and about the incoherent bands.

\subsection{ARPES from a layered system\label{sec:2D}}

Many systems of current interest
are built of weakly coupled layers or chains: High-Tc cuprates
and Fe-based superconductors, quasi-one-dimensional magnetic systems,
ruthenites, iridates, etc. If the surface coincides 
with the two-dimensional layer or is built of one-dimensional (1D) 
chains we can neglect the dispersion in the surface-normal direction. Then
the planes become decoupled, and the GF
of Eq.~(\ref{eq:Gl1l2a1a2}), does not depend on $l_{1}$ and $l_{2}$. For 
the Mott-Hubbard system we can write 
$G_{\mathbf{k}_{\parallel},l_{1},l_{2},\alpha_{1},\alpha_{2}}(\omega)
=\delta_{l_{1},l_{2}}G_{b,\mathbf{k}_{\parallel}}(\omega)$.
Equation (\ref{eq:Gl1l2calG}) then yields 
\begin{align*}
\mathcal{G}\left(\hat{q},\omega\right) & =\left\langle \!\left\langle 
\hat{C}_{\sigma}|\hat{C}_{\sigma}^{\dagger}\right\rangle \!
\right\rangle _{\omega}\propto G_{b,\mathbf{k}_{\parallel}}(\omega),\\
\mathcal{A}\left(\hat{q},\omega\right) & \propto A_{b}\left(\mathbf{k},
\omega\right)
\end{align*}
Thus, for systems with a negligible dispersion normal to the 
surface ARPES directly measures the SF of the electron GF.

\subsection{Account of the surface in a 3D system}

However, the actual crystals are three-dimensional. 
Even in quasi-1D or quasi-2D systems 
the chains or the layers are coupled,
and the energy of the photohole disperses with $k_{\perp}$. This
dispersion may be small compared with the dispersion parallel
to the surface, but with the progress in angular and energy
resolution \cite{Borisenko12} it becomes measurable, 
which calls for a more thorough theoretical
analysis of the surface-normal degree of freedom, which is proposed below. 

In the equation of motion for the GF 
\begin{equation*}
\omega\left\langle \!\left\langle a_{\mathbf{k}_{1}}|a_{\mathbf{k_{2}}}^{\dagger}\right\rangle 
\!\right\rangle _{\omega}=\delta_{\mathbf{k_{1},k}_{2}}+\left(\varepsilon_{\mathbf{k}}
+\Sigma_{\mathbf{k},\omega}\right)\left\langle \!\left\langle a_{\mathbf{k}_{1}}|
a_{\mathbf{k_{2}}}^{\dagger}\right\rangle \!\right\rangle _{\omega},
%\label{eq:eom1D}
\end{equation*}
which straightforwardly follows from Eqs. (\ref{eq:aadagg}) and (\ref{eq:Gkkb}),
we perform in both sides the Fourier transform 
$a_{\mathbf{k}_{\parallel},l}=\left(1/\sqrt{N_{\perp}}\right)\sum_{p}\mathrm{e}^{iplc}a_{\mathbf{k}}$
inverse to (\ref{eq:ak}) and obtain the equation of motion for 
the ``interlayer'' GF of Eq.~(\ref{eq:Gl1l2a1a2}) 
\begin{align}
\omega G_{\mathbf{k}_{\parallel},l_{1},l_{2}}\left(\omega\right) & 
=\delta_{l_{1},l_{2}}+\sum_{l}h_{l_{1},l}(\mathbf{k}_{\parallel},\omega)
G_{\mathbf{k}_{\parallel},l,l_{2}}\left(\omega\right),\label{eq:moz}\\
h_{l,l_{2}}(\mathbf{k}_{\parallel},\omega) & \equiv\frac{1}{N_{\perp}}
\sum_{p}e^{ip\left(l-l_{2}\right)c}\left(\varepsilon_{\mathbf{k}}
+\Sigma_{\mathbf{k},\omega}\right).\label{eq:hzz2}
\end{align}
Equation (\ref{eq:moz}) has the form of an equation of motion
for an effective 1D tight-binding system with an
energy-dependent (and generally non-Hermitian) Hamiltonian 
\begin{equation}
\hat{h}(\mathbf{k}_{\parallel},\omega)=\sum_{l_{1},l_{2}}h_{l_{1},l_{2}}(
\mathbf{k}_{\parallel},\omega)a_{\mathbf{k}_{\parallel},l_{1}}^{\dagger}
a_{\mathbf{k}_{\parallel},l_{2}}\label{eq:hzz}
\end{equation}
with hopping amplitudes given by Eq. (\ref{eq:hzz2}).

Now we proceed with a semi-infinite crystal. The surface may be introduced
as a perturbation $\hat{V}$ that breaks an infinite crystal into
two non-interacting parts. In Refs.\ \onlinecite{Kalkstein71,Liebsch03}
the coupling between the two parts is eliminated by means of a non-diagonal
perturbation $V_{l,l_{2}}=-h_{l,l_{2}}$. We achieve the same result
using the diagonal perturbation of the form 
\begin{equation}
\hat{V}=\varepsilon_{0}\sum_{i}a_{\mathbf{k}_{\parallel},i}^{\dagger}
a_{\mathbf{k}_{\parallel},i},\label{eq:Vs}
\end{equation}
where $i$ enumerates the atomic planes of a slab that divides the
crystal into two semi-infinite parts. The width of the slab should
be equal or larger than the maximal distance $\left(l-l_{2}\right)c$
for which the hopping integrals $h_{l,l_{2}}$ in Eq.~(\ref{eq:moz})
are non-zero. In the limit $\varepsilon_{0}\rightarrow\infty$ the
two half-spaces are separated by an infinite barrier. Similar approaches
are used for the description of vacancies \cite{Economou}, in the
cavity method of DMFT \cite{Georges96}, and for the hard-core constraint
for magnon pairs in acute-angle helimagnets \cite{Kuzian07,Nishimoto15}
(the bound states of magnons being analogues of the surface states).

Note that the perturbation $\hat{V}$ leads to a relaxation of the
system, which changes the effective Hamiltonian $\hat{h}(\mathbf{k}_{\parallel},\omega)$.
These changes are expected to be localized at the surface and, in
principle, can be taken into account in a self-consistent way. 
Here we neglect it and consider the simplest case when only 
adjacent planes are coupled by $\hat{h}(\mathbf{k}_{\parallel},\omega)$
\begin{align}
\varepsilon_{\mathbf{k}} & =\varepsilon_{\mathbf{k}_{\parallel}}
-2t_{\mathbf{k}_{\parallel}}\cos pc,\label{eq:epsk}\\
\Sigma_{\mathbf{k},\omega} & =\Sigma_{\mathbf{k}_{\parallel},\omega}
-2\tau_{\mathbf{k}_{\parallel},\omega}\cos pc,\label{eq:sgmk}
\end{align}
$c$ being the inter-plane distance. Then we may retain in Eq. (\ref{eq:Vs})
only the term with $z_{i}=0$. The assumption (\ref{eq:epsk}) is
natural for a narrow-band system, and the local character of the self-energy
(\ref{eq:sgmk}) is also a commonly accepted approximation 
\cite{Georges96,Kotliar06,Kuchinskii12}.
Note that we do not make any assumptions about the $\mathbf{k}_{\parallel}$-
dependence of the self-energy, which may be quite strong
\cite{Kuzian98,Kuzian99,Hayn00,Kuchinskii12}. We then obtain the bulk 
GF of Eq.~(\ref{eq:Gkkb}) in the form 
\begin{equation}
G_{b,\mathbf{k}}(\omega)=\left[\omega-\sigma_{\mathbf{k}_{\parallel},\omega}
+2T_{\mathbf{k}_{\parallel},\omega}\cos pc\right]^{-1},\label{eq:Gbulk}
\end{equation}
where we have included the dispersion parallel to the surface $\varepsilon_{\mathbf{k}_{\parallel}}$
into the real part of the self-energy: $\sigma_{\mathbf{k}_{\parallel},
\omega}\equiv\varepsilon_{\mathbf{k}_{\parallel}}+\Sigma_{\mathbf{k}_{\parallel},\omega}$
and $T_{\mathbf{k}_{\parallel},\omega}\equiv t_{\mathbf{k}_{\parallel}}
+\tau_{\mathbf{k}_{\parallel},\omega}$.

The equation of motion for the GF of the perturbed system then reads
\begin{align}
\omega G_{\mathbf{k}_{\parallel},l_{1},l_{2}} & =\delta_{l_{1},l_{2}}
+\left(\varepsilon_{\mathbf{k}_{\parallel}}+\Sigma_{\mathbf{k}_{\parallel},
\omega}\right)G_{\mathbf{k}_{\parallel},l_{1},l_{2}}\nonumber \\
 & -T_{\mathbf{k}_{\parallel},\omega}\left(G_{\mathbf{k}_{\parallel},
 l_{1}+1,l_{2}}+G_{\mathbf{k}_{\parallel},l_{1}-1,l_{2}}\right)\nonumber \\
 & +\delta_{l_{1},0}\varepsilon_{0}G_{\mathbf{k}_{\parallel},0,l_{2}},
 \label{eq:omGkapzz}
\end{align}
We perform the double Fourier transform 
\begin{equation*}
G_{\mathbf{k}_{\parallel},p_{1},p_{2}}(\omega)=\frac{1}{N_{\perp}}
\sum_{l,l^{\prime}}e^{-i\left(p_{1}l+ip_{2}l^{\prime}\right)c}
G_{\mathbf{k}_{\parallel},l,l^{\prime}}(\omega),
%\label{eq:2Fourier}
\end{equation*}
in both sides of Eq. (\ref{eq:omGkapzz}) to obtain for the GF 
of Eq.~(\ref{eq:Gp1p2a1a2}) 
\begin{equation}
G_{\mathbf{k}_{\parallel},p_{1},p_{2}}(\omega)=G_{b,\mathbf{k}_{1}}(\omega)
\left(\delta_{p_{1},p_{2}}+\frac{\varepsilon_{0}}{\sqrt{N_{\perp}}}
\mathsf{G}_{0}\right),\label{eq:Gkappp}
\end{equation}
where we have defined 
\begin{equation}
\mathsf{G}_{l}\equiv G_{\mathbf{k}_{\parallel},l,p_{2}}=\frac{1}{\sqrt{N_{\perp}}}
\sum_{p}e^{iplc}G_{\mathbf{k}_{\parallel},p,p_{2}}.\label{eq:G0}
\end{equation}
Now we substitute (\ref{eq:Gkappp}) into the left hand side of Eq.
(\ref{eq:G0}) for $l=0$ and find 
\begin{align}
\mathsf{G}_{0} & =\frac{1}{\sqrt{N_{\perp}}}\frac{G_{b,\mathbf{k_{2}}}(
\omega)}{1-\varepsilon_{0}g_{\mathbf{k}_{\parallel}}(\omega)},\nonumber \\
g_{\mathbf{k}_{\parallel}}(\omega) & \equiv\frac{1}{N_{\perp}}
\sum_{p}G_{b,\mathbf{k}}(\omega).\label{eq:gkapdef}
\end{align}
Finally, Eq. (\ref{eq:Gkappp}) gives the GF of the perturbed system
\begin{align}
G_{\mathbf{k}_{\parallel},p_{1},p_{2}}(\omega) & =G_{b,\mathbf{k}_{2}}(
\omega)\left\{ \delta_{p_{1},p_{2}}+\frac{\varepsilon_{0}G_{b,\mathbf{k}_{1}}(
\omega)}{N_{\perp}\left[1-\varepsilon_{0}g_{\mathbf{k}_{\parallel}}(
\omega)\right]}\right\} \label{eq:Geps0}\\
 & \xrightarrow[\varepsilon_{0}\rightarrow\infty]{}G_{b,\mathbf{k}_{2}}(
 \omega)\left\{ \delta_{p_{1},p_{2}}-\frac{G_{b,\mathbf{k}_{1}}(\omega)}{N_{\perp}
 g_{\mathbf{k}_{\parallel}}(\omega)}\right\} .\label{eq:Gs}
\end{align}
Thus, we have found the GF of the Hamiltonian 
$\hat{h}_{1}(\mathbf{k}_{\parallel},\omega)=\hat{h}(\mathbf{k}_{\parallel},\omega)
+\varepsilon_{0}a_{\mathbf{k}_{\parallel},0}^{\dagger}a_{\mathbf{k}_{\parallel},0}$.
Equation (\ref{eq:Gs}) is the desired result: it gives the GF for
the semi-infinite crystal that extends over the half-space $z\leq z_{0}=-c$,
which is necessary for the calculation of ARPES via Eqs.~(\ref{eq:jCCdag})
and (\ref{eq:Gp1p2calG}).

Note that the approximations given by Eqs. (\ref{eq:epsk}) and (\ref{eq:sgmk})
may be easily relaxed by using a thicker slab in Eq.~(\ref{eq:Vs}).
In this case, the GF may be found by successively applying this trick
\cite{Economou,Kuzian07,Nishimoto15}: based on Eq. (\ref{eq:Geps0})
we find the GF of the Hamiltonian $\hat{h}_{2}(\mathbf{k}_{\parallel},\omega)
=\hat{h}_{0}(\mathbf{k}_{\parallel},\omega)+\varepsilon_{0}a_{
\mathbf{k}_{\parallel},c}^{\dagger}a_{\mathbf{k}_{\parallel},c}$
with two perturbed planes and employ it to find the GF for three perturbed
planes, etc.

A similar technique may be used to account for the surface relaxation
of the system. In this case, the charge self-consistency may require
the diagonal terms $h_{l,l}(\mathbf{k}_{\parallel},\omega)$ (\ref{eq:hzz2})
to be $l$-dependent \cite{Potthoff96,Potthoff99}, and also the non-diagonal
terms $h_{l,l_{2}}(\mathbf{k}_{\parallel},\omega)$ of the effective
Hamiltonian (\ref{eq:hzz}) may depend on both indices $l$ and $l_{2}$
rather than on their difference. These deviations from the Hamiltonian
(\ref{eq:hzz}) obtained from the bulk values of 
$\varepsilon_{\mathbf{k}}+\Sigma_{\mathbf{k},\omega}$
are expected to have local character, and, thus, can be treated by
Eqs. (\ref{eq:omGkapzz}) - (\ref{eq:Geps0}). Thereby, the problem
is reduced to the problem of a few impurities in a 1D chain. These
changes will perturb the electronic structure near the surface, and
surface states may emerge. As mentioned above, the surface states
that decouple from the bulk continuum have close analogy to the bound
states of magnons in 1D magnets \cite{Kuzian07}.

\section{ARPES from a 3D Mott-Hubbard system\label{sec:3D}}

In this section we consider the case when one of the waves dominates
the sum in Eq.~(\ref{eq:fim}), so the time-reversed LEED function
$\varphi_{>}^{*}\left(\mathbf{x},\hat{q},E\right)$ (\ref{eq:LEED})
inside the solid may be approximated by a single evanescent wave:
\begin{equation}
\varphi_{>}^{*}\left(\mathbf{x},\hat{q},E\right)\approx e^{i\left[
\mathbf{k}_{\parallel}\mathbf{x}_{\parallel}+k_{\perp}\left(z-z_{0}
\right)\right]}u\left(\mathbf{x},\mathbf{k}_{\parallel},E\right),
\label{eq:LEEDmfp}
\end{equation}
where $k_{\perp}=k_{z}^{\prime}-ik_{z}^{\prime\prime}$. 
Then Eq.~(\ref{eq:Gp1p2calG}) acquires the form
\begin{align}
\mathcal{G}\left(\hat{q},\omega\right) & \equiv\left\langle \!\left
\langle \hat{C}\left(\hat{q},E\right)|\hat{C}^{\dagger}
\left(\hat{q},E\right)\right\rangle \!\right\rangle _{\omega}=\left|
\mathsf{M}\left(\mathbf{k}_{\parallel},E\right)\right|^{2}\times\nonumber \\
 & \times\frac{N_{\parallel}}{N_{\perp}}\sum_{p_{1},p_{2}}e^{i(p_{1}
 -p_{2})z_{0}}\Delta_{p_{1}}\Delta_{p_{2}}^{\ast}G_{\mathbf{k}_{\parallel},
 p_{1},p_{2}}(\omega),\label{eq:GMH}
\end{align}
where we define $\omega\equiv E-\hbar\Omega$. 
Substituting the expression (\ref{eq:Gs}) for the GF of the semi-infinite
system into Eq.~(\ref{eq:GMH}) and setting there $z_{0}=-c$ (the
surface layer) we have 
\begin{equation}
\mathcal{G}\left(\hat{q},\omega\right)=\left|\mathsf{M}\left(
\mathbf{k}_{\parallel},E\right)\right|^{2}N_{\parallel}\left[I_{1}
\left(\mathbf{k}_{\parallel},\omega\right)-I_{2}\left(\mathbf{k}_{\parallel},
\omega\right)\right],\label{eq:GI1I2}
\end{equation}
where 
\begin{align}
I_{1}\left(\mathbf{k}_{\parallel},\omega\right) & 
=\frac{1}{2\pi c}\intop_{-\pi/c}^{\pi/c}\left|\Delta_{p}\right|^{2}G_{b,
\mathbf{k}}(\omega)dp,\label{eq:I1}\\
I_{2}\left(\mathbf{k}_{\parallel},\omega\right) & =\frac{1}{g_{\mathbf{k}_{
\parallel}}(\omega)}I_{21}(k_{z}^{\prime})I_{21}(-k_{z}^{\prime}),\label{eq:I2}\\
I_{21}(k_{z}^{\prime}) & =\frac{1}{2\pi c}\intop_{-\pi/c}^{\pi/c}
\Delta_{p}e^{-ipc}G_{b,\mathbf{k}}(\omega)dp.\label{eq:I21}
\end{align}
Here $G_{b,\mathbf{k}}(\omega)$ is given by Eq. (\ref{eq:Gbulk})
with $\mathbf{k}=\mathbf{k}_{\parallel}+p\mathbf{n}$, 
$\Delta_{p}=\left\{ 1-\exp{\left[i\left(k_{z}^{\prime}-p\right)-
k_{z}^{\prime\prime}\right]c}\right\} ^{-1}$.
The integrand of $I_{1}$ is defined by the bulk GF of Eq.~(\ref{eq:Gkkb}),
and $I_{2}$ comes from the surface term of (\ref{eq:Gs}).

The integrals (\ref{eq:I1}) and (\ref{eq:I21}) are calculated using
the residue theorem (see the details in Appendix) 
\begin{align}
I & =\intop_{-\pi}^{\pi}R(\cos\varphi,\sin\varphi)d\varphi\nonumber \\
 & =\oint_{|z|=1}R_{0}(z)dz=2\pi i\sum_{m=1}^{n}\mathrm{Res}_{z=z_{m}}R_{0}(z),
 \label{eq:Res}
\end{align}
where $R(u,v)$ is a rational function of $u$ and $v$, and $z_{m}$,
$m=1,\ldots,n$ are poles of rational function $R_{0}(z)=
-\frac{i}{z}R\left[\frac{1}{2}\left(z+\frac{1}{z}\right),
\frac{1}{2}\left(z-\frac{1}{z}\right)\right]$
that lie inside the circle $|z|<1$. We have two residues for $I_{1}$,
\begin{equation}
I_{1}\left(\mathbf{k}_{\parallel},\omega\right)=R_{1}\left(
\mathbf{k}_{\parallel},\omega\right)+R_{2}\left(\mathbf{k}_{\parallel},
\omega\right),\label{eq:I1R1R2}
\end{equation}
which are given by Eqs. (\ref{eq:R1}) and (\ref{eq:R2}) and one
residue for $I_{21}$ (\ref{eq:I2kap}). The final expression is 
\begin{equation}
\mathcal{G}\left(\hat{q},\omega\right)=K\left(\mathbf{k}_{\parallel},
E\right)F\left(\mathbf{k},\omega-\sigma_{\mathbf{k}_{\parallel},\omega}\right),
\label{eq:Gfin}
\end{equation}
where 
\begin{align}
K\left(\mathbf{k}_{\parallel},E\right) & =\frac{\left|\mathsf{M}
\left(\mathbf{k}_{\parallel},E\right)\right|^{2}N_{\parallel}}{1-
e^{-2k^{\prime\prime}c}}, \nonumber \\
%\label{eq:Kkw}\\
F\left(\mathbf{k},\epsilon\right) & =\frac{1}{\epsilon-\epsilon_{\mathbf{k}
,\omega}-B_{\mathbf{k},\omega}^{2}g_{s}(\epsilon,T_{\mathbf{k}_{\parallel},\omega})},
\label{eq:Fke}\\
\epsilon_{\mathbf{k},\omega} & \equiv-2T_{\mathbf{k}_{\parallel},\omega}
e^{-k_{z}^{\prime\prime}c}\cos k_{z}^{\prime}c, \nonumber \\
%\label{eq:ekw}\\
B_{\mathbf{k},\omega}^{2} & \equiv T_{\mathbf{k}_{\parallel},\omega}^{2}
\left(1-e^{-2k_{z}^{\prime\prime}c}\right), \nonumber
%\label{eq:Bkw}
\end{align}
and the function 
\begin{align}
g_{s}(\epsilon,b) & \equiv\left\langle \!\left\langle a_{0}|
a_{0}^{\dagger}\right\rangle \!\right\rangle _{\epsilon} \nonumber \\
%\label{eq:tau2a}\\
 & =1/\left\{ \epsilon-b^{2}/\left[\epsilon-b^{2}/\left(\epsilon-
 \cdots\right)\right]\right\} \label{eq:taucf}\\
 & =1/\left(\epsilon-b^{2}g_{s}(\epsilon,b)\right)\\
 & =\left\{ \epsilon-\mathrm{sgn}\left[\mathrm{Re}\left(\epsilon\right)
 \right]\sqrt{\epsilon^{2}-4b^{2}}\right\} /2b^{2}\label{eq:tauw}
\end{align}
is the GF for the states localized at the edge of a semi-infinite
chain described by the tight-binding Hamiltonian \cite{Haydock80,Henk93}
$\hat{h}=b\sum_{l\geq0}a_{l}^{\dagger}a_{l+1}$.

Equation (\ref{eq:taucf}) represents the function $F\left(\mathbf{k},\epsilon\right)$
in a continued-fraction form. This ensures the correct analytic properties
of $\mathcal{G}\left(\hat{q},\omega\right)$, see Eq.~(\ref{eq:GI1I2}),
as a function of complex energy $\epsilon=\omega-\sigma_{\mathbf{k}_{\parallel},
\omega}=\omega-\varepsilon_{\mathbf{k}_{\parallel}}-\Sigma_{\mathbf{k}_{\parallel},\omega}$.
The GF is an analytic function in the whole complex energy plane with
the exception of the real axis, where it may have poles and branch
cuts \cite{Haydock80}. In the upper (lower) half-plane it coincides
with the retarded (advanced) GF. It is easy to see that the function
$F\left(\mathbf{k},\omega-\sigma_{\mathbf{k}_{\parallel},\omega}\right)$
coincides with the bulk GF of Eq.~(\ref{eq:Gbulk}) in the so-called 
direct-transitions limit $k_{z}^{\prime\prime}\rightarrow0$. In this limit it is 
$\epsilon_{\mathbf{k},\omega}\rightarrow-2T_{\mathbf{k}_{\parallel},\omega}\cos k_{z}^{\prime}c$
and $B_{\mathbf{k},\omega}^{2}\rightarrow0$, and finally Eq.~(\ref{eq:Fke})
becomes 
\begin{equation*}
F\left(\mathbf{k},\omega\right)\xrightarrow[k_{z}^{\prime\prime}\to 0]{}
G_{b,\mathbf{k}_{\parallel}+k_{z}^{\prime}\mathbf{n}}(\omega).
%\label{eq:dirtr}
\end{equation*}
A pole of the bulk GF $G_{b,\mathbf{k}_{\parallel}+k_{z}^{\prime}\mathbf{n}}(\omega)$
may occur if both $\Sigma_{\mathbf{k},\omega}$ is real and the energy
$\omega_{0}\left(\mathbf{k}\right)$ satisfies the equation 
\begin{equation}
\omega_{0}\left(\mathbf{k}\right)=\sigma_{\mathbf{k}_{\parallel},\omega_{0}}
-2T_{\mathbf{k}_{\parallel},\omega_{0}}\cos k_{z}^{\prime}c.\label{eq:w0}
\end{equation}
In the vicinity of this energy it is $G_{b,\mathbf{k}_{\parallel}
+k_{z}^{\prime}\mathbf{n}}(\omega)\approx Z_{\mathbf{k}}(\omega_{0})/(\omega
-\omega_{0})$
with the residue $Z_{\mathbf{k}}(\omega)=\left[1-\partial
\Sigma_{\mathbf{k},\omega}/\partial\omega\right]^{-1}$.
For $k_{z}^{\prime\prime}c\ll1$ the pole transforms into a resonance
of a Lorentzian form 
\begin{align}
F_{r} & \approx\frac{Z_{\mathbf{k}}\left(\omega_{r}\right)\left(1
+k_{z}^{\prime\prime}c\right)}{2k_{z}^{\prime\prime}c\left[\omega
-\omega_{r}+i\Gamma\right]},\label{eq:Fr}\\
\Gamma & \approx2Z_{\mathbf{k}}\left(\omega_{r}\right)\left|
T_{\mathbf{k}_{\parallel},\omega_{r}}\right|k_{z}^{\prime\prime}c
\sqrt{\left(k_{z}^{\prime\prime}c\right)^{2}+\sin^{2}k_{z}^{\prime}c}
\label{eq:Gfull}\\
 & \approx k_{z}^{\prime\prime}v_{h}, \quad
v_{h} \equiv 2Z_{\mathbf{k}}\left(\omega_{r}\right)\left|T_{\mathbf{k}_{
\parallel},\omega_{r}}\sin k_{z}^{\prime}c\right|c, \label{eq:Gvh}
%\label{eq:72}
\end{align}
where the energy of the resonance satisfies the equation 
$\omega_{r}-\sigma_{\mathbf{k}_{\parallel},\omega_{r}}=-2T_{
\mathbf{k}_{\parallel},\omega_{r}}\cos k_{z}^{\prime}c/\cosh 
k_{z}^{\prime\prime}c$,
which for a small decay index $k_{z}^{\prime\prime}c$ gives
$\omega_{r}\left(\mathbf{k}\right)\approx\omega_{0}\left(\mathbf{k}\right)$
Equation~(\ref{eq:Gvh}) is
the well-known expression for the resonance width 
$\Gamma$ \cite{Starnberg93,Smith93,Krasovskii07}
in terms of the the group velocity of the hole quasi-particle 
$v_{h}=\partial\omega_{0}/\partial k_{z}^{\prime}$.
Formula (\ref{eq:Gfull}) shows that this expression
is valid only in the middle of the quasi-particle band, where 
$k_{z}^{\prime}\gg k_{z}^{\prime\prime}$.
The expression for $F_{r}$ for arbitrary values of $k_{z}^{\prime\prime}c$
is given in the Appendix (\ref{eq:Frfull}).

The energy dependence of the final state (\ref{eq:LEEDmfp}) leads
to the energy dependence of the GF of Eq.~(\ref{eq:Gfin}) 
via the functions
$K\left(\mathbf{k}_{\parallel},E\right)$, $k_{z}^{\prime}\left(\mathbf{k}_{\parallel},E\right)$,
and $k_{z}^{\prime\prime}\left(\mathbf{k}_{\parallel},E\right)$.
Now let us assume that the matrix element slowly varies with energy 
\begin{equation*}
\left|\mathsf{M}\left(\mathbf{k}_{\parallel},E\right)\right|^{2}
\approx\left|\mathsf{M}\left(\mathbf{k}_{\parallel},E_{0}\right)\right|^{2}
\end{equation*}
and concentrate on the role of the decay of the final states into the solid.
For the analysis of photoemission in the next section we introduce
the normalized GF 
\begin{align}
\tilde{\mathcal{G}}\left(\hat{q},\omega\right) & =
\frac{\mathcal{G}(\hat{q},\omega)}{K\left(\mathbf{k}_{\parallel},
E_{0}\right)}\label{eq:Gnorm}\\
 & =\frac{K\left(\mathbf{k}_{\parallel},E\right)}{K\left(
 \mathbf{k}_{\parallel},E_{0}\right)}F\left(\mathbf{k},\omega-
 \sigma_{\mathbf{k}_{\parallel},\omega}\right)
\end{align}
and its SF 
\begin{align}
A(\omega) & =-\frac{1}{\pi}\mathrm{Im}\mathcal{\tilde{G}}\left(\hat{q},
\omega\right)=A_{1}(\omega)-A_{2}(\omega),\label{eq:A}\\
A_{1}(\omega) & =\left(1-e^{-2k_{0}^{\prime\prime}c}\right)\left(
-\frac{1}{\pi}\mathrm{Im}I_{1}\right),\label{eq:A1}\\
A_{2}(\omega) & =\left(1-e^{-2k_{0}^{\prime\prime}c}\right)\left(
-\frac{1}{\pi}\mathrm{Im}I_{2}\right).\label{eq:A2}
\end{align}
Here $A_{1}$ and $A_{2}$ give the contributions to the photocurrent
from the bulk and the surface terms of the GF of Eq.~(\ref{eq:Gs}),
respectively. This normalization facilitates the comparison with the
bulk SF $A_{\mathrm{b}}(\mathbf{k},\omega)$, which is normalized
to unity.

\section{Discussion\label{sec:Discussion}}

In this section, we study the behavior of the GF (\ref{eq:Gnorm}) and the 
relevant SFs (\ref{eq:A})--(\ref{eq:A2}), which define the shape
of the photocurrent energy distribution curve (EDC) through the expression
(\ref{eq:ACCdag}). 
We neglect the $\omega $-dependence of the effective hopping in the normal direction, 
 and in Eq.~(\ref{eq:Gbulk}) we set $T_{\mathbf{k}_{\parallel},\omega}\equiv T$. We assume 
that $T$ is real and positive, and take it as the unit of energy. We chose 
the value of the parallel momentum $\mathbf{k}_{\parallel}$ that gives
$\varepsilon_{\mathbf{k}_{\parallel}}=-2.2T$, in order to have the quasiparticle peaks close to the 
lower Hubbard band for our model of the initial state self-energy, see 
see next subsection.  This choice  highlights
the quasiparticle band narrowing and the 
dispersion of the lower Hubbard band weight in the surface-normal direction.

\subsection{Initial states}\label{susec:i}

\begin{figure}[htb]
\includegraphics[width=1\columnwidth]{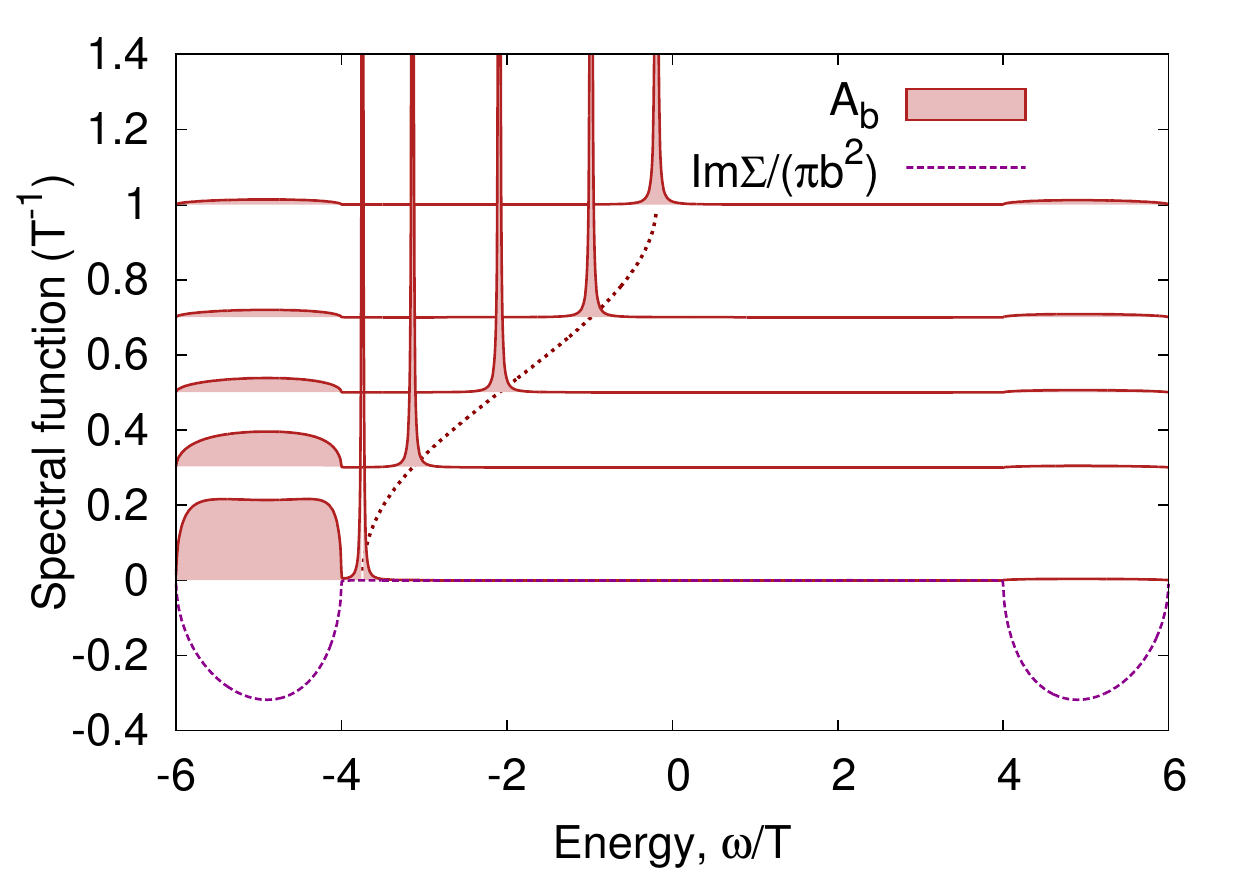} 
\caption{(Color online) The spectral density of the bulk Green's function (\ref{eq:Gbulk})
$A_{\mathrm{b}}(\mathbf{k},\omega+i\eta)=-\mathrm{Im}G_{b,\mathbf{k}}(\omega+i\eta)/\pi$
(\ref{eq:Ab}) for $\varepsilon _{\mathbf{k}_{\parallel}}=-2.2T$,
$k_{z}^{\prime}c=0$, $0.3\pi$, $0.5\pi$, $0.7\pi$, $\pi$ (from
bottom to the top), and the imaginary part of its self-energy 
$\mathrm{Im}\Sigma_{\mathbf{k}_{\parallel},\omega+i\eta}/\pi b^{2}$
(\ref{eq:Sgmodel});
$b_{1}=5T$, $b_{2}=b=T$.
Here and below a small imaginary constant $i\eta,\eta=0.001$ is added
to the energy argument in order to visualize the coherent $\delta$-function
peaks in the spectral densities. Dark red dotted line shows the quasi-particle
dispersion 
$k_{z}=\arccos\left[(\sigma_{\mathbf{k}_{\parallel},\omega}-\omega)/2T\right]$. }

\label{fb} 
\end{figure}
\begin{figure}[htb]
\includegraphics[width=1\columnwidth]{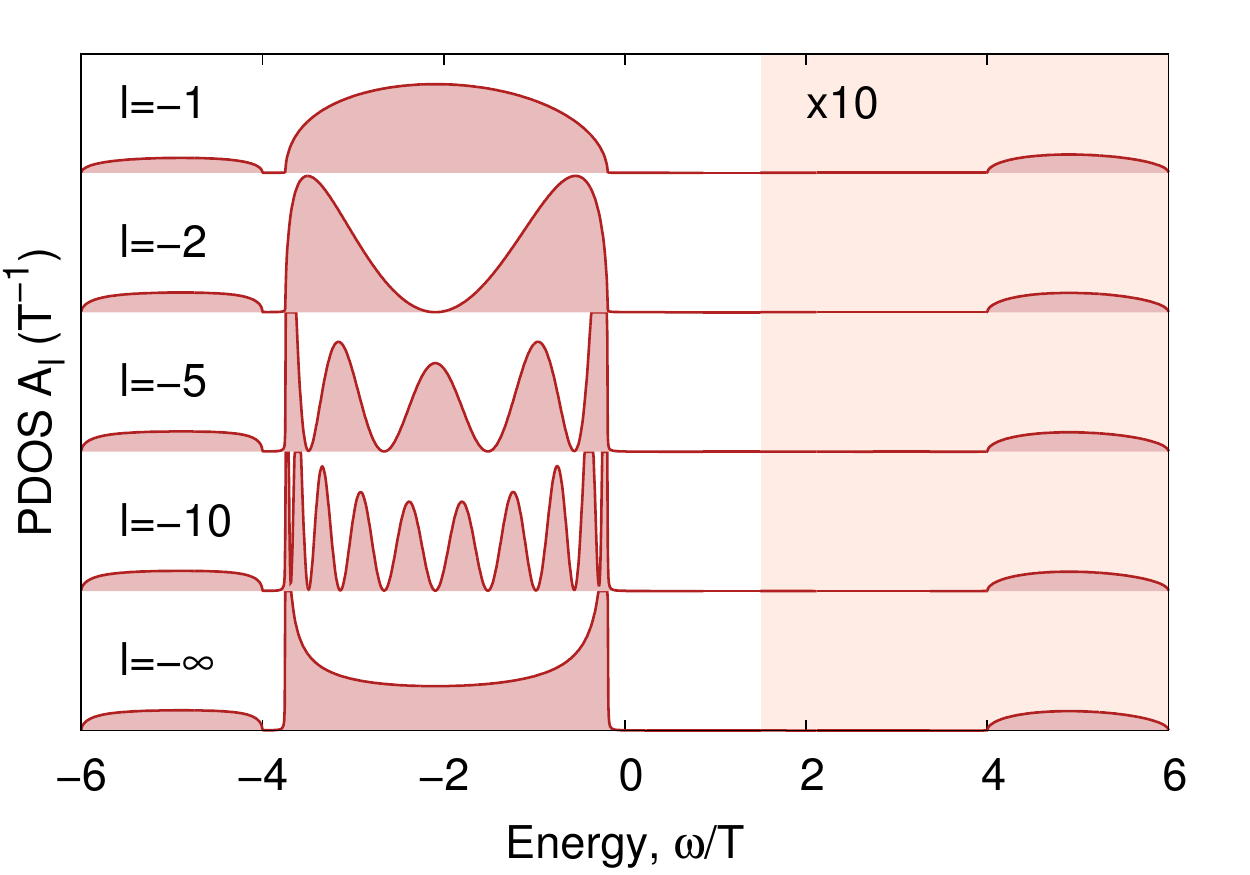} 
\caption{(Color online) The density of states $A_{l}$ (\ref{eq:All}) 
projected on a Bloch sum of Wannier functions located in the layer $z=lc$, 
Eq. (\ref{eq:fikapl}).
$\varepsilon _{\mathbf{k}_{\parallel}}=-2.2T$. 
}
\label{pdos} 
\end{figure}

We adopt a simple analytic expression for the self-energy of the bulk
GF, see Eq.~(\ref{eq:sgmk}): 
\begin{align}
\Sigma_{\mathbf{k}_{\parallel},\omega} & =b^{2}t_{2b}(\omega),\label{eq:Sgmodel}\\
t_{2b}(\omega) & =1/\left\{ \omega-b_{1}^{2}/\left[\omega-b_{2}^{2}/
\left(\omega-b_{1}^{2}t_{2b}(\omega)\right)\right]\right\} \nonumber \\
%\label{eq:t2bcf}\\
 & =\frac{\omega^{2}-b_{1}^{2}+b_{2}^{2}+s_{t}\sqrt{\left(\omega^{2}
 -b_{1}^{2}-b_{2}^{2}\right)^{2}-4b_{1}^{2}b_{2}^{2}}}{2\omega b_{2}^{2}},\nonumber
 %\label{eq:t2bw}
\end{align}
where $s_{t}(\omega)=-\mathrm{sgn}\left[\mathrm{Re}\left(\omega^{2}-b_{1}^{2}-b_{2}^{2}\right)\right]$.
By choosing $b_{1}>b_{2}$ and an appropriate value for $b$ we construct
the function $G_{b,\mathbf{k}}(\omega)$, Eq.~(\ref{eq:Gbulk}),
with a three-peak structure of the SF characteristic of a Mott-Hubbard
metal \cite{Byczuk07}. 
It has a central coherent peak at $\omega=\omega_{0}(\mathbf{k})$, 
see Eq.~(\ref{eq:w0}), and two incoherent bands over
the energy intervals $\left(b_{1}-b_{2}\right)^{2}<\omega^{2}<
\left(b_{1}+b_{2}\right)^{2}$.
Figure\ \ref{fb} shows the SF (\ref{eq:Ab}) for several values
of $k_{z}$ and 
$b_{1}=5T$, $b_{2}=b=T$.
Note that the quasi-particle band is narrower than in the non-interacting
case $\Sigma_{\mathbf{k}_{\parallel},\omega}=0$, where its width
is $4T$. This follows from Eq.~(\ref{eq:w0})
because within the Hubbard gap the self-energy is approximately
$\mathrm{Re}\left(\Sigma_{\mathbf{k},\omega}\right)\sim-\alpha \omega$,
with a positive coefficient $\alpha $ being weakly dependent
on $\omega$
(see, e.g., Fig.~2c of Ref.~\onlinecite{Byczuk07}). Then Eq.~(\ref{eq:w0})
gives a renormalization of the dispersion 
$\omega_{0}\left(\mathbf{k}\right)\sim\varepsilon_{\mathbf{k}}/(1+\alpha )$.

The incoherent bands originate from the self-energy branch cuts, where
its imaginary part is a negative definite function, see the dashed
line in Fig.~\ref{fb}. Its position does not depend on $k_{z}$
because we have chosen the coefficients $b_{1}$ and $b_{2}$ to be
$k_{z}$-independent. Nevertheless, its intensity is pronouncedly
momentum dependent. This dependence has the same origin
as the quasiparticle band narrowing. It comes from
the spectral weight redistribution, which is the consequence of the
coupling between the quasi-particle and the incoherent bands. 
To show this, we note that the SF obeys the sum rule 
\begin{equation*}
\int_{-\infty}^{\infty}\omega A_{\mathrm{b}}(\mathbf{k},\omega+i0)d\omega
=\varepsilon_{\mathbf{k}}.
%\label{eq:epsmf}
\end{equation*}
At a fixed momentum $\mathbf{k}$, the spectral density has one coherent
peak situated at $\omega_{0}(\mathbf{k})$ between two incoherent
bands 
\begin{align*}
A_{\mathrm{b}}(\mathbf{k},\omega+i0) & =Z_{\mathbf{k}}(\omega_{0})\delta(\omega
-\omega_{0})+A_{\mathrm{inc}}(\mathbf{k},\omega),\\
A_{\mathrm{inc}}(\mathbf{k},\omega) & =A_{\mathrm{lhb}}(\mathbf{k},\omega)
+A_{\mathrm{uhb}}(\mathbf{k},\omega),
\end{align*}
where $A_{\mathrm{lhb}}(\mathbf{k},\omega)$ and $A_{\mathrm{uhb}}(\mathbf{k},\omega)$
are the low and the upper Hubbard band SFs, respectively, that yield
the humps at the energies $\omega_{\mathrm{lhb/uhb}}\sim\pm b_{1}$.
Then, for the incoherent bands we obtain 
\begin{align}
\int_{-\infty}^{\infty}\omega A_{\mathrm{inc}}(\omega)d\omega & 
\sim\omega_{\mathrm{lhb}}W_{\mathrm{lhb}}(\mathbf{k})+
\omega_{\mathrm{uhb}}W_{\mathrm{uhb}}(\mathbf{k})\label{eq:Winc}\\
 & \sim\omega_{0}(\mathbf{k})\left(1+\alpha -Z_{\mathbf{k}}\right).
\end{align}
The spectral weights $W_{\mathrm{lhb}}(\mathbf{k})$ and
$W_{\mathrm{uhb}}(\mathbf{k})$, obviously depend on $\mathbf{k}$.
In Eq.~(\ref{eq:Winc}), we approximated the
integrals $\int_{-\infty}^{\infty}\omega A_{\mathrm{i}}(\omega)d\omega$
by $\omega_{\mathrm{i}}W_{\mathrm{i}}(\mathbf{k})$, 
where ``i'' is ``lhb'' or ``uhb''.
The momentum dependence of the
incoherent band weight was recently observed in ARPES experiments
and in the DMFT calculations for vanadates Sr$_{x}$Ca$_{1-x}$VO$_{3}$
\cite{Takizawa09,Yoshida10}.

It is instructive to calculate the DOS projected on the 2D Bloch sum
of Wannier functions, Eq.~(\ref{eq:fikapl}), in the layer $z=lc$: 
\begin{align}
A_{l} & =-\frac{1}{\pi}\mathrm{Im}G_{\mathbf{k}_{\parallel},l,l}(\omega),
\label{eq:All}\\
G_{\mathbf{k}_{\parallel},l,l} & \equiv\left\langle \!\left
\langle a_{\mathbf{k}_{\parallel},l}|a_{\mathbf{k}_{\parallel},l}^{\dagger}
\right\rangle \!\right\rangle _{\omega}\nonumber \\
 & =\frac{1}{N_{\perp}}\sum_{p_{1},p_{2}}e^{i(p_{1}-p_{2})lc}
 G_{\mathbf{k}_{\parallel},p_{1},p_{2}}(\omega),\nonumber 
\end{align}
where $G_{\mathbf{k}_{\parallel},p_{1},p_{2}}(\omega)$ is the GF
of the semi-infinite system given by Eq.~(\ref{eq:Gs}). The integrals
over $p_{1}$ and $p_{2}$ are calculated using Eq. (\ref{eq:Res}),
and the result looks very simple: 
\[
G_{\mathbf{k}_{\parallel},l,l}=g_{\mathbf{k}_{\parallel}}
\left(\omega\right)\left\{ 1-\left[Tg_{s}
\left(\omega-\sigma_{\mathbf{k}_{\parallel},\omega},T
\right)\right]^{2\left|l\right|}\right\} ,
\]
where $g_{s}(\omega)$ is given by Eq.~(\ref{eq:tauw}). The function
$g_{\mathbf{k}_{\parallel}}\left(\omega\right)$, Eqs. (\ref{eq:gkapdef})
and (\ref{eq:gkap}), is the bulk value of the layer function 
$G_{\mathbf{k}_{\parallel},l,l}$,
$l\to\infty$. Figure \ref{pdos} shows the result for layers at different
depths. We see that the projected DOS has rather peculiar dependence
on $l$ (cf. Sec.~4 and Fig.~1 of Ref. \onlinecite{Henk93}),
and its convergence to the bulk shape is slow. Strong oscillations
of the DOS near the surface were also documented in \textit{ab initio}
calculations, see, e.g., Fig.~4(d) in Ref.~\onlinecite{Rybkin12}.

\subsection{Final states}

\begin{figure}[htb]
\includegraphics[width=1\columnwidth]{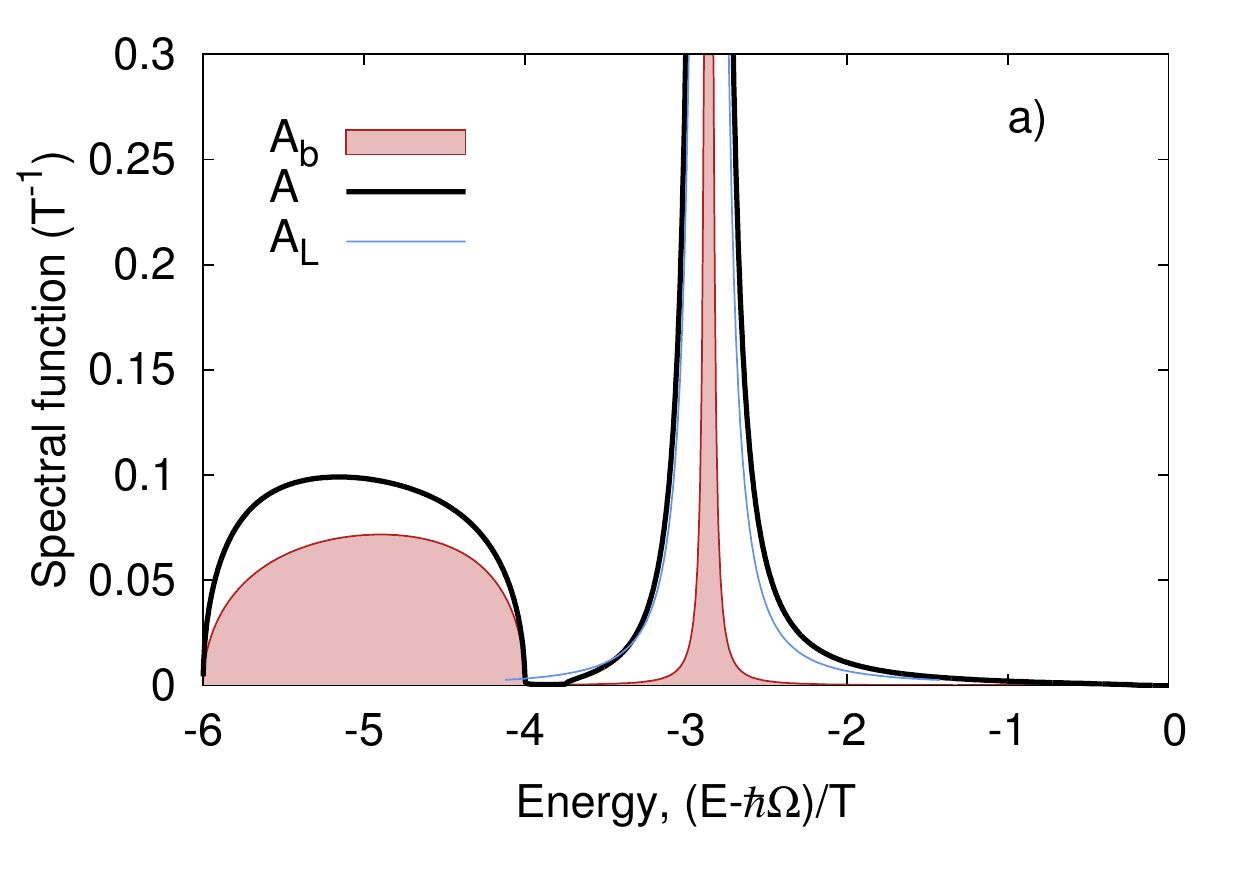}\\
 \includegraphics[width=1\columnwidth]{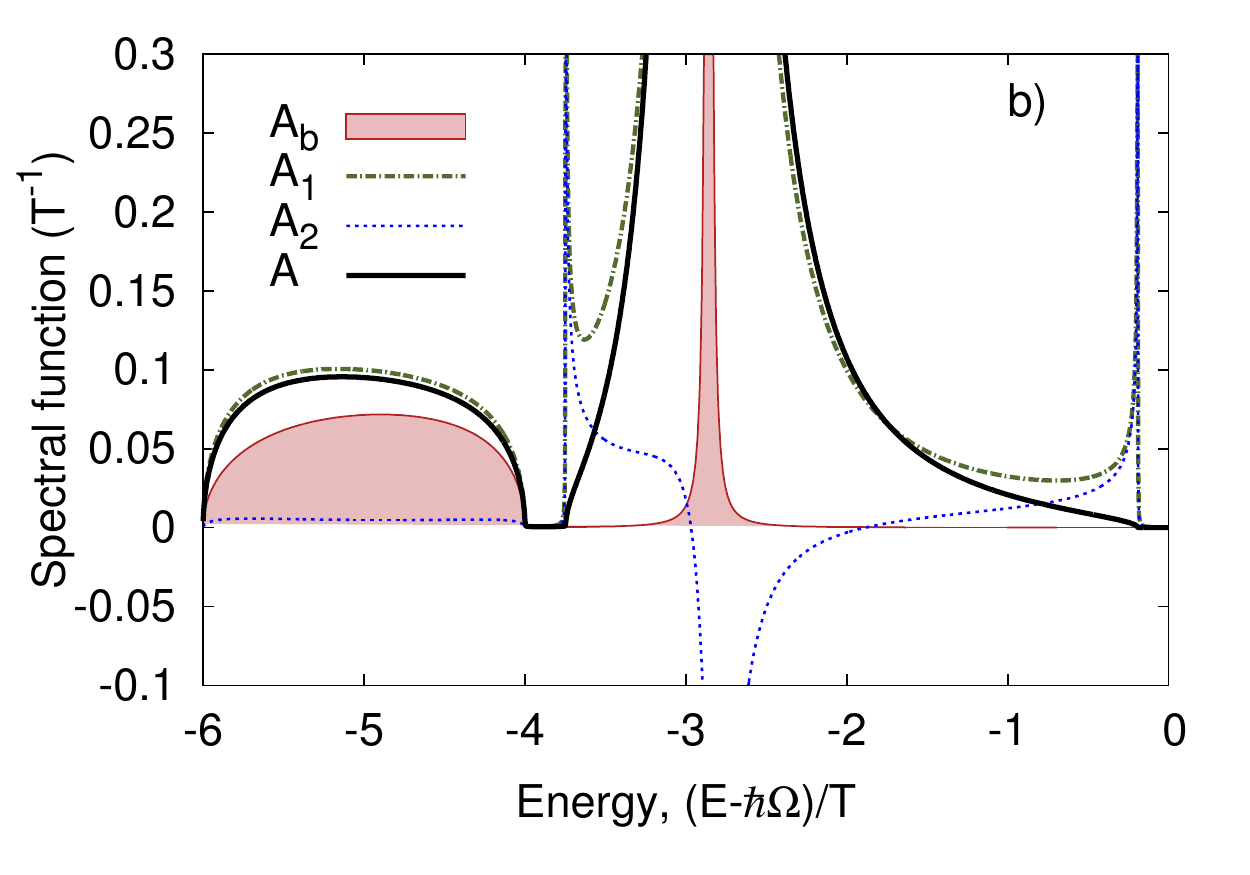} 
 \caption{(Color online) The spectral density $A(\omega+i\eta)$ (\ref{eq:A})
compared with the spectral density $A_{\mathrm{b}}(\mathbf{k},\omega+i\eta)$
(\ref{eq:Ab}) of the bulk Green's function (\ref{eq:Gbulk}) shown
in Fig.\ \ref{fb} for (a) $V_{\rm i}=0.1T$ and (b) $V_{\rm i}=T$.
The parameters are
$\epsilon(\mathbf{k}_{0})=\hbar\Omega-2.86T$,
$\varepsilon_{\mathbf{k}_{\parallel}}=-2.2T$
$\epsilon_{z}^{\prime}=10T/c$, $k_{0}c=0.36\pi$.
The contributions coming from the bulk [$A_{1}(\omega+i\eta)$,
Eq. (\ref{eq:A1})], and the surface [$A_{2}(\omega+i\eta)$,
Eq. (\ref{eq:A2})] terms of the Green's function (\ref{eq:Gs})
are also shown. Thin cornflower blue line in the upper panel shows
the spectral density behavior near the ($\delta$-functional) coherent
peak of the bulk Green's function Lorentzian (\ref{eq:Fr}) }

\label{far} 
\end{figure}
\begin{figure}[htb]
\includegraphics[width=1\columnwidth]{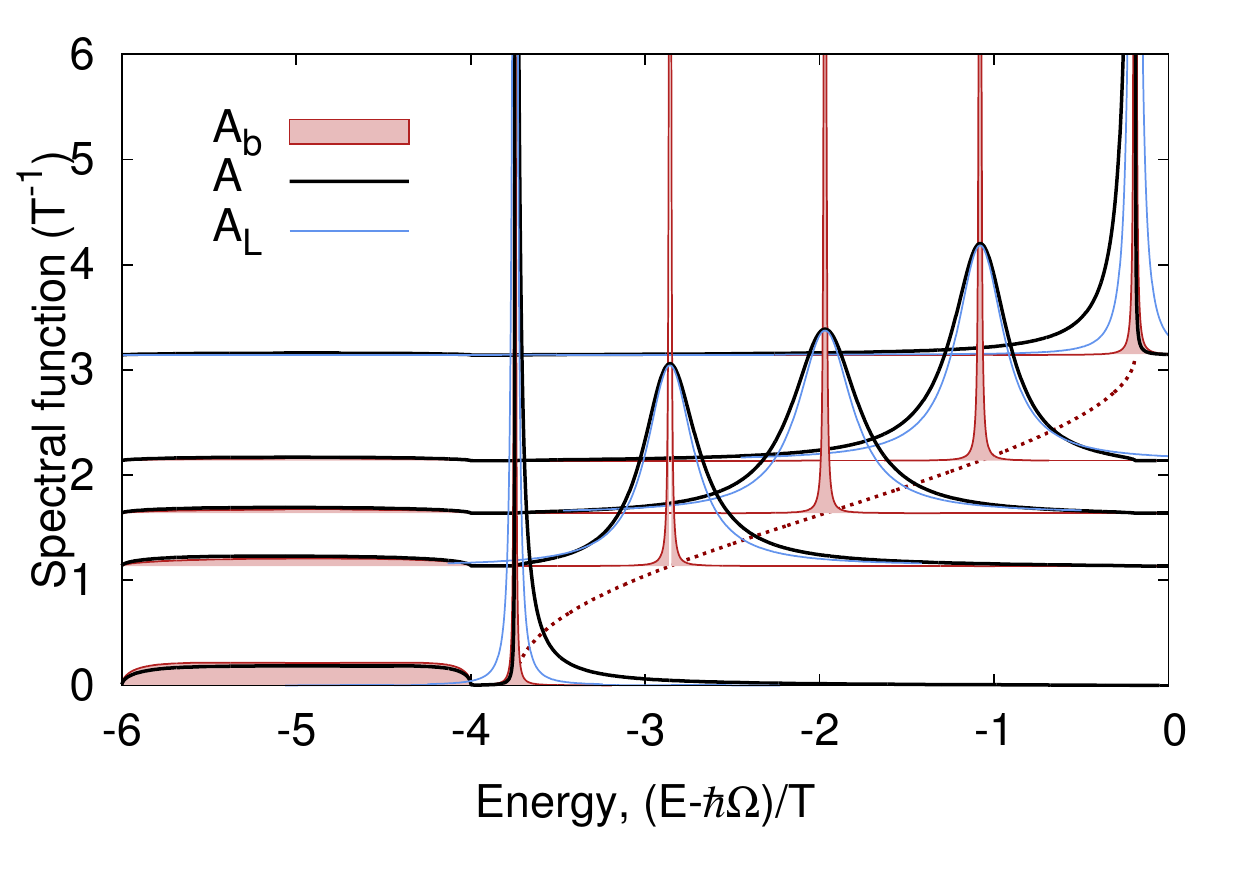} 
\caption{(Color online) The spectral density $A(\omega+i\eta)$ (\ref{eq:A})
for different photon energies $\hbar\Omega$. The lowest 
curve corresponds to $k_{0}c=0$, and the topmost one to $k_{0}c=\pi$. 
The energy step between the curves is $\hbar\Delta\Omega=0.89T$ and
$V_{\rm i}=T$. The notation is the same as in the previous figures.}
\label{fark} 
\end{figure}

In order to take into account the inelastic scattering of electrons
in the LEED experiment Slater\cite{Slater37} proposed to add an imaginary
term to the potential energy. The Schr\"odinger equation then reads
\begin{equation}
\left[-\frac{1}{2}\nabla^{2}+V(\mathbf{x})-iV_{\rm i}\right]\varphi_{>}
\left(\mathbf{x}\right)=E\varphi_{>}\left(\mathbf{x}\right),
\label{eq:SchrEq}
\end{equation}
where $V(\mathbf{x})$ is the periodic potential inside the solid.
The optical potential $V_{{\rm i}}$ may be considered an approximation
for the imaginary part of the electron self-energy.

Following Ref.~\onlinecite{Slater49}, we start from the solution of the 
unperturbed problem 
\begin{equation*}
\left[-\frac{1}{2}\nabla^{2}+V(\mathbf{x})\right]\varphi_{0}
\left(\mathbf{k},\mathbf{x}\right)=\epsilon(\mathbf{k})\varphi_{0}
\left(\mathbf{k},\mathbf{x}\right)
\end{equation*}
in Wannier representation 
\begin{equation*}
\varphi_{0}\left(\mathbf{k},\mathbf{x}\right)=\frac{1}{\sqrt{
N_{\parallel}N_{\perp}}}\sum_{\mathbf{R}}\mathrm{e}^{i\mathbf{k}
\mathbf{R}}w_{f}\left(\mathbf{x}-\mathbf{R}\right)
\end{equation*}
and search the solution of the perturbed problem (\ref{eq:SchrEq}) in the form 
\[
\varphi_{>}\left(\mathbf{x}\right)=\sum_{\mathbf{R}}\Psi(\mathbf{R})
w_{f}\left(\mathbf{x}-\mathbf{R}\right).
\]
Then the modulating function $\Psi(\mathbf{R})$ is the solution of
the equation 
\begin{equation}
\left[\hat{\epsilon }\left(-i\nabla\right)-iV_{\rm i}\right]\Psi\left(\mathbf{x}\right)
=E\Psi\left(\mathbf{x}\right).\label{eq:efSchrEq}
\end{equation}
Here $\hat{\epsilon }\left(-i\nabla\right)$ is a differential operator 
obtained from the function $\epsilon({\mathbf k})$ by the substitution 
${\mathbf k}\to -i\nabla$.
Thus, Eq.~(\ref{eq:efSchrEq}) is a Schr\"odinger equation for 
$\Psi\left(\mathbf{x}\right)$,
in which the perturbation $-iV_{\rm i}$ is the potential 
energy, while the kinetic energy operator is derived from the 
band structure $\epsilon(\mathbf{k})$ of the unperturbed problem.

Substituting $\Psi(\mathbf{x})=\exp\left(i\mathbf{k}\mathbf{x}\right)$
with complex $\mathbf{k}=\mathbf{k}_{\parallel}-\left(k_{z}^{\prime}
+ik_{z}^{\prime\prime}\right)\mathbf{n}$
(in LEED, $\mathbf{k}_{\perp}$ points into the crystal),
we obtain 
\begin{equation}
\epsilon\left[\mathbf{k}_{\parallel}-\left(k_{z}^{\prime}
+ik_{z}^{\prime\prime}\right)\mathbf{n}\right]=E+iV_{{\rm i}},
\label{eq:V2k}
\end{equation}
which allows to find the
components of complex $\mathbf{k}$-vector from the 
analytical continuation of the function 
$\epsilon(\mathbf{k})$ into the complex energy plane.

\subsection{Final state energy far from the gap\label{sub:fargap}}

When the energies of photoelectrons
$E$ are far from the gaps in the unoccupied
spectrum $\epsilon(\mathbf{k})$, we may write 
\begin{align*}
\epsilon\left[\mathbf{k}_{\parallel}-\left(k_{z}^{\prime}
+ik_{z}^{\prime\prime}\right)\mathbf{n}\right] &
 \approx\epsilon(\mathbf{k}_{0})+\epsilon_{z}^{\prime}\left(\delta
 +ik_{z}^{\prime\prime}\right) \\
 %\label{eq:Efar}\\
\mathbf{k}_{0} & \equiv\mathbf{k}_{\parallel}-k_{0}\mathbf{n},\nonumber \\
\epsilon_{z}^{\prime} & \equiv-\frac{\partial\epsilon\left(\mathbf{k}
\right)}{\partial k_{z}}\vert_{\mathbf{k}=\mathbf{k}_{0}}>0,\nonumber \\
\delta & \equiv k_{z}^{\prime}-k_{0}.\nonumber 
\end{align*}
Equation~(\ref{eq:V2k}) then gives $k_{z}$ with a constant imaginary
part and a real part that linearly depends on energy 
\begin{align}
k_{z}^{\prime} & =k_{0}+\frac{E-\epsilon(\mathbf{k}_{0})}{
\epsilon_{z}^{\prime}(\mathbf{k}_{0})},\label{eq:kpfar}\\
k_{z}^{\prime\prime} & =\frac{V_{i}}{\epsilon_{z}^{\prime}(\mathbf{k}_{0})}.
\label{eq:kppfar}
\end{align}

Figures \ref{far} and \ref{fark} show typical EDCs in this regime
in comparison with the bulk SF. The coherent peaks, which are $\delta$-functions
in our approximation for the initial states, transform into Lorentzians,
whose width according to Eq.~(\ref{eq:Gvh}) is proportional to $k_{z}^{\prime\prime}$
and to the quasi-particle group velocity perpendicular to the surface
\cite{Starnberg93,Smith93}. The broadening is invisible for the incoherent
part, but the dependence of the real part of the wave vector $k_{z}^{\prime}$
on energy (\ref{eq:kpfar}) leads to deviations of the EDC shape from
the SF as a result of the dispersion of the intensity of the incoherent
part with $k_{z}^{\prime}$. 
\begin{figure*}[htb]
\includegraphics[width=0.29\textwidth]{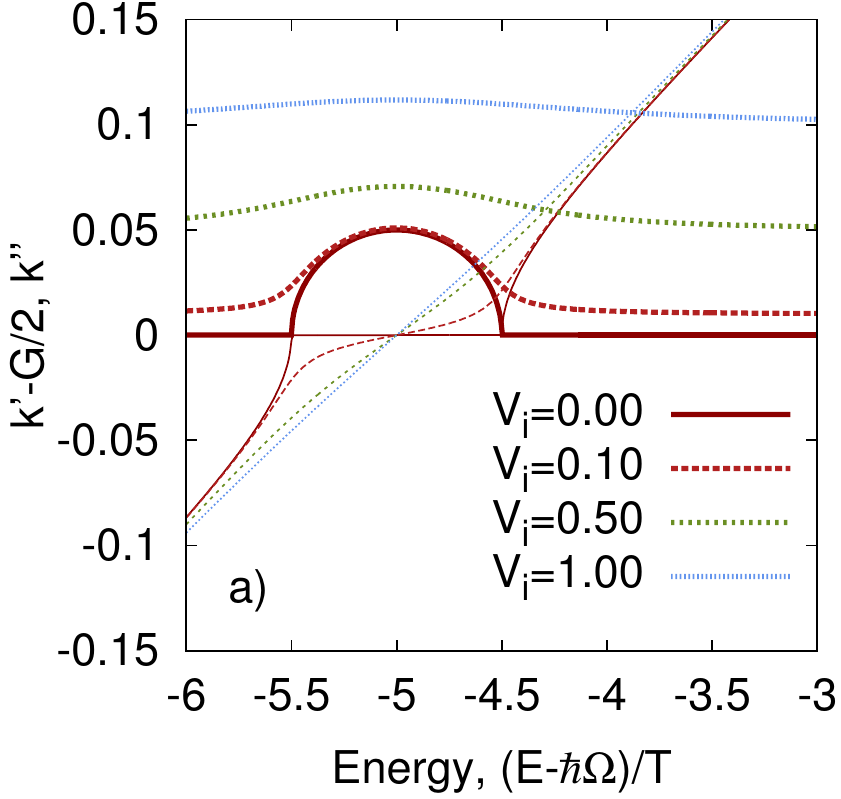} 
\includegraphics[width=0.3\textwidth]{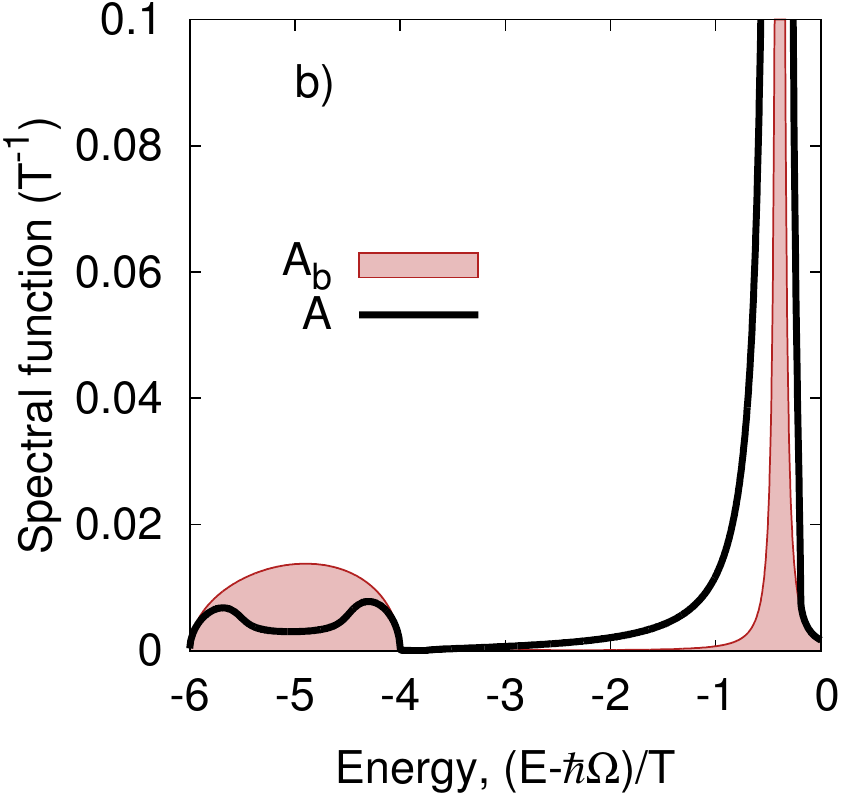}
\includegraphics[width=0.3\textwidth]{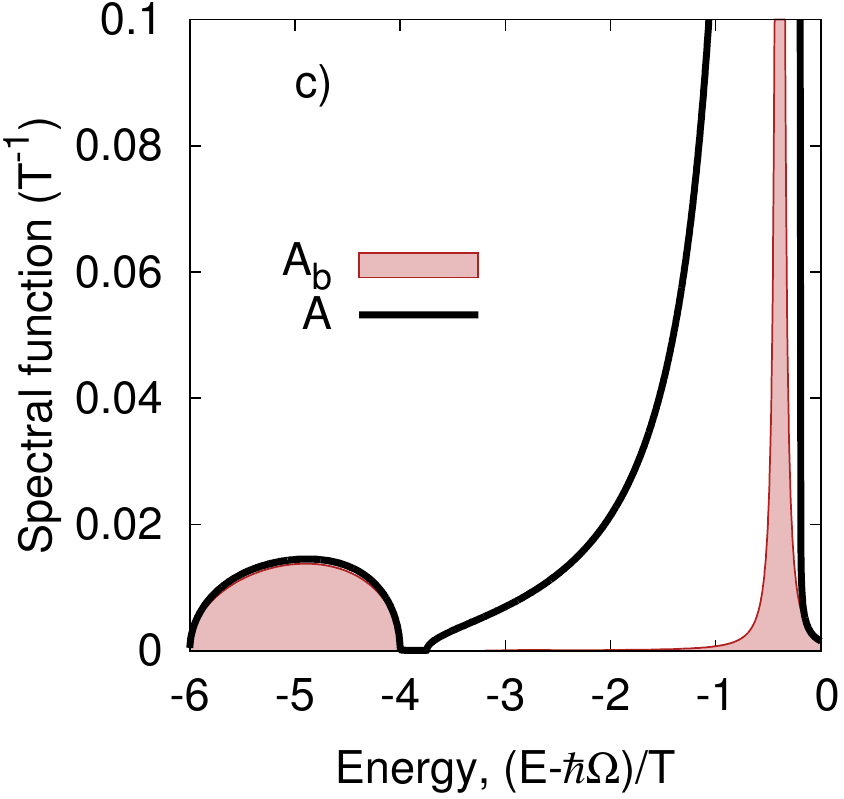} 
\caption{(Color online) \textbf{ a) } $k_{z}^{\prime}$ (\ref{eq:kpgap}) -
thin lines, and $k_{z}^{\prime\prime}$ (\ref{eq:kppgap}) - thick
lines for various values of ``optic potential'' $V_{i}$; 
$E_{G}=\hbar\Omega-5T$,
$\varepsilon_{\mathbf{k}_{\parallel}}=-2.2T$,
$W=0.5T$, $\epsilon_{z}^{\prime}=10T/c$.
The spectral density $A(\omega+i\eta)$ compared with the spectral
density of the bulk Green's function (\ref{eq:Gbulk}) shown in Fig.\ \ref{fb}
for \textbf{ b) } $V_{i}=0.1T$, and \textbf{ c)
} $V_{i}=T$. The energy $E_{0}=\hbar\Omega$
is used for the normalization in Eqs. (\ref{eq:A})-(\ref{eq:A2})}

\label{f} 
\end{figure*}

\subsection{Final states near the Bragg gap}

Now we consider the case when the energy of photoelectrons is close
to a gap in the spectrum $\epsilon(\mathbf{k})$, i.e. 
$\mathbf{k}=\mathbf{k}_{\parallel}+\left(\frac{G}{2}+\delta\right)\mathbf{n}$
is near a Brillouin zone boundary, $G\mathbf{n}$ being a reciprocal
lattice vector. Then $\epsilon(\mathbf{k})$ can be approximated as
\begin{equation*}
\epsilon(\mathbf{k})\approx E_{G}\pm\sqrt{W^{2}+\left(\epsilon_{z}^{
\prime}\right)^{2}\delta^{2}},
%\label{eq:Egap}
\end{equation*}
where $E_{G}$ and $W$ are the center and the half-width of the gap,
and $\epsilon_{z}^{\prime}$ is some positive value, which plays the
role of a ``bare group velocity'' in the absence of coupling between
waves with $\mathbf{k}$ and $\mathbf{k}+G\mathbf{n}$. Equation~(\ref{eq:V2k})
now gives 
\begin{align}
k_{z}^{\prime} & =\frac{G}{2}+\delta,\label{eq:kpgap}\\
\delta & =\mathrm{sgn}(\Delta E)\frac{\sqrt{R(\Delta E)+\left(\Delta E-V^{2}
-W^{2}\right)}}{\epsilon_{z}^{\prime}},\nonumber \\
k_{z}^{\prime\prime} & =\frac{\sqrt{R(\Delta E)-\left(\Delta E-V^{2}
-W^{2}\right)}}{\epsilon_{z}^{\prime}},\label{eq:kppgap}\\
\Delta E & \equiv E-E_{G},\nonumber \\
R(\Delta E) & \equiv\sqrt{\left(\Delta E-V^{2}-W^{2}\right)^{2}
+4V^{2}\left(\Delta E\right)^{2}}.\nonumber 
\end{align}
Figure \ref{f}(a) shows that both real and imaginary parts of $k_{z}$
become energy dependent. The dependences (\ref{eq:kpgap}) and (\ref{eq:kppgap})
are smoothed by the optical potential $V_{{\rm i}}$. Figure \ref{f}(b)
shows that for small values of $V_{{\rm i}}$ the EDC from the incoherent
band may substantially deviate from the SF. This may be important
for the interpretation of the low-energy ARPES, where the inelastic
scattering is relatively weak.

\subsection{Small inelastic mean free path.}

It is instructive to consider the limit $k_{z}^{\prime\prime}c\to-\infty$,
in which case the final state is strongly localized near
the surface. Let us consider the bulk contribution $I_{1}$ to the
GF, see Eq. (\ref{eq:GI1I2}). In this limit, the only non-vanishing
part is $R_{2}$, see Eqs. (\ref{eq:I1R1R2}) and (\ref{eq:R2}),
which yields 
\begin{equation*}
I_{1}=\frac{S_{\omega}}{\sqrt{\left(\omega-\sigma_{\mathbf{k}_{\parallel},
\omega}\right)^{2}-4T^{2}}},
%\label{eq:I1loc}
\end{equation*}
where 
\begin{equation}
S_{\omega}\equiv\mathrm{sgn}\left[\mathrm{Re}\left(\omega-\sigma_{
\mathbf{k}_{\parallel},\omega}\right)\right].\label{eq:Sw}
\end{equation}
For the coherent band, the self-energy is real. Then the SF $A_{1}$,
which is proportional to the imaginary part of $I_{1}$, see Eq. (\ref{eq:A1}),
is nonzero for $\left(\omega-\sigma_{\mathbf{k}_{\parallel},\omega}\right)^{2}
-4T^{2}<0$.
Near the points $\omega-\sigma_{\mathbf{k}_{\parallel},\omega}=\pm2T$,
the bulk contribution $A_{1}$ to the SF has horn-like singularities
typical of the 1D density of states (cf. the lowest curve in Fig.
\ref{pdos}). These ``horns'' are cancelled by the same singularities
in the surface term, see Eq. (\ref{eq:I2}): 
\begin{equation*}
I_{2}=\frac{S_{\omega}\left[\omega-\sigma_{\mathbf{k}_{\parallel},\omega}
-S_{\omega}\sqrt{\left(\omega-\sigma_{\mathbf{k}_{\parallel},\omega}\right)^{2}
-4T^{2}}\right]^{2}}{4T^{2}\sqrt{\left(\omega-\sigma_{\mathbf{k}_{\parallel},\omega}\right)^{2}
-4T^{2}}}.
%\label{eq:I2loc}
\end{equation*}
Substituting both expressions into (\ref{eq:GI1I2}), we obtain 
\begin{align}
 & \mathcal{G}\left(\hat{q},\omega\right)\propto\left(I_{1}
 -I_{2}\right)\nonumber \\
 & =\frac{\omega-\sigma_{\mathbf{k}_{\parallel},\omega}
 -S_{\omega}\sqrt{\left(\omega-\sigma_{\mathbf{k}_{\parallel},\omega}
 \right)^{2}-4T^{2}}}{2T^{2}}.\label{eq:Gse}
\end{align}
It is clear that the imaginary part of this expression as a function
of $\omega$ has the semi-elliptical form of the local DOS at the
edge site of a \emph{semi-infinite} chain (cf. the topmost curve in
Fig. \ref{pdos}). Equation (\ref{eq:Gse})
follows from the general formula (\ref{eq:Gfin}) in the limit 
$e^{-k_{z}^{\prime\prime}c}\to0$.
Figure~\ref{far}(b) demonstrates
that a similar cancellation happens also for finite $k_{z}^{\prime\prime}$.
Thus, the account of surface terms in the initial state GF of Eq.~(\ref{eq:Gs})
is crucial for the coherent contribution but less important for the
incoherent band.

\section{Concluding remarks\label{sec:Concluding-remarks}}

In strongly correlated systems, the conventional understanding of
the solid as a Fermi-liquid of quasiparticles breaks down in the sense
that a considerable part of the spectral weight transfers from the
quasiparticle peak to the incoherent band. This occurs because the
removal of an electron from an $N$-electron state creates a superposition
of $(N-1)$-electron eigenstates with a spread of energies. Thus, the
electronic structure of strongly correlated system is described by
a momentum-dependent spectral function $A_{\mathrm{b}}(\mathbf{k},\omega)$
(\ref{eq:Ab}) rather than a quasiparticle energy $\varepsilon_{\mathbf{k}}$.
The incoherent bands in the  spectral function
of the ground state come from the correlated motion of electrons
expressed as the imaginary part of the self-energy. In ARPES,
these bands are observed as structureless humps apart from pronounced
quasiparticle peaks.

ARPES data provide the information about both initial and final states
of the photoemission process. Both kinds of states characterize the
solid under study, and they are solutions of the Schr\"odinger
equation with the same Hamiltonian. Using the sudden approximation,
we have shown how the spectra depend on physical properties of the
initial and final states. First, we recast the well-known mean-field
theory expression for the photocurrent in the one-step approach as
a formula for a DOS-function projected onto a surface-localized electron
state $\chi$, Eq.~(\ref{eq:chiC}). The wave function $\chi(\mathbf{r})$
decays into the solid owing to the spatial
decay of the time-reversed LEED function, and, at the same time, it
rapidly vanishes in the vacuum owing to the confinement of the initial
states. Then the many-body calculation of the ARPES intensity
in the one-step approach reduces to the calculation of a
spectral function of the two-time retarded GF for an operator that
creates an electron in the state $\chi$.

Further, we make use of the Wannier representation and obtain the
GF for a semi-infinite system out of the GF of an infinite system.
This approach is especially advantageous for strongly correlated
systems. For the simplest case of a one-band Mott-Hubbard
system and neglecting the modification of the crystal potential at
the surface we have obtained an analytical result. Combined
with modern numerical methods, our approach is fully applicable to
realistic models of surfaces.

Furthermore, for the present model we have obtained an analytical
expression for the photocurrent assuming that the inelastic scattering
in the final state can be described by a mean free path. Here we approximated
the LEED function inside the solid by a single evanescent wave. This
is not a serious limitation,
which may be easily lifted within the present formalism. Expressions
(\ref{eq:ACCdag}) and (\ref{eq:Gfin}) explicitly relate the energy
distribution of the photocurrent to the \emph{bulk} electronic structure.

The analysis of the expressions reveal the following features of the
photocurrent: (i) As in the mean-field theory \cite{Starnberg93,Smith93,Krasovskii07},
the quasi-particle pole of the bulk Green's functions 
gives rise to a resonance, whose width is proportional to the imaginary
part of the wave vector $k_{z}^{\prime\prime}$ and to the group velocity
of the hole perpendicular to the surface. (ii) For the incoherent
band, even if its energy range is $\mathbf{k}$-independent, as is
the case in most DMFT theories, its spectral weight turns out momentum-dependent
\cite{Takizawa09,Yoshida10}. Apart form the obvious $\mathbf{k}_{\parallel}$-dependence
of the intensity, this manifests itself in the photon energy dependence
of the EDC. This reflects in the first place the $\mathbf{k}$-dependence
of the initial-state spectral function $A_{\mathrm{b}}(\mathbf{k},\omega)$,
Eq.~(\ref{eq:Ab}), but it may also involve more complicated 
matrix element effects. This happens already in the simplest case when the 
final-state decay rate $k_{z}^{\prime\prime}$ is constant over the whole EDC
energy range. However, when the energy passes through a gap in the
final-state spectrum, where $k_{z}^{\prime\prime}$ rapidly changes
with energy, EDC becomes dramatically distorted with respect to the
underlying spectral function $A_{\mathrm{inc}}(\mathbf{k},\omega)$.
Furthermore, interesting interference effects
are expected when the LEED function has several evanescent components
with different decay rates \cite{Krasovskii07b}.

In this work, we have considered rather simple models of both initial
and final states. The present formalism
opens a way to study ARPES for
more realistic models of strongly
correlated electron systems, such as LDA+DMFT
\cite{Kotliar06,Nekrasov06}, LDA + Gutzwiller method \cite{Deng08,Ho08,Deng09},
LDA-based many-band Hubbard model \cite{Kuzian12,Monney13},
embedded-cluster quantum chemistry calculations \cite{Katukuri14,Nishimoto16},
etc. In addition, a more accurate
treatment of final-state effects can be implemented \cite{Krasovskii07,Krasovskii07b,Cui10}.
Owing to the simplicity of the Wannier representation,
the present formalism can be straightforwardly extended to two-photon
and pump-probe photoemission \cite{Schattke08,Braun15}. 
\begin{acknowledgments}
NATO (Belgium), Grant No. SfP-984735 is acknowledged with gratitude.
This work was supported by the Spanish Ministry of Economy and Competitiveness
MINECO (Project No. FIS2013-48286-C2-1-P). 
\end{acknowledgments}
\appendix
%dummy comment inserted by tex2lyx to ensure that this paragraph is not empty
%dummy comment inserted by tex2lyx to ensure that this paragraph is not empty
%dummy comment inserted by tex2lyx to ensure that this paragraph is not empty
%\global\long\def\thefigure{A\arabic{figure}}\setcounter{figure}{0}

\section{Details of derivation}

\label{AppA}\begin{widetext}

Let us consider the double integral in Eq.(\ref{eq:j2}) 
\begin{align}
I & =\iint d^{3}\mathbf{x}_{1}d^{3}\mathbf{x}_{2}\left[\varphi_{1}^{\prime}
+i\varphi_{1}^{\prime\prime}\right]\hat{O}_{1}G_{1,2}^{\prime\prime}\hat{O}_{2}
\left[\varphi_{2}^{\prime}-i\varphi_{2}^{\prime\prime}\right]\nonumber \\
 & =\iint_{\mathbf{x}_{1,2}\subset\mathcal{S}}d^{3}\mathbf{x}_{1}d^{3}
 \mathbf{x}_{2}\left[\varphi_{1}^{\prime}\hat{O}_{1}G_{1,2}^{\prime\prime}
 \hat{O}_{2}\varphi_{2}^{\prime}+\varphi_{1}^{\prime\prime}\hat{O}_{1}
 G_{1,2}^{\prime\prime}\hat{O}_{2}\varphi_{2}^{\prime\prime}+i\left(
 \varphi_{1}^{\prime\prime}\hat{O}_{1}G_{1,2}^{\prime\prime}\hat{O}_{2}
 \varphi_{2}^{\prime}-\varphi_{1}^{\prime}\hat{O}_{1}G_{1,2}^{\prime\prime}
 \hat{O}_{2}\varphi_{2}^{\prime\prime}\right)\right]\label{eq:I}
\end{align}
where we use the simplified notations $\varphi_{i}\equiv\varphi_{>}
\left(\mathbf{x}_{i},\hat{q},E\right)$,
$\hat{O}_{i}\equiv\hat{O}\left(\mathbf{x}_{i}\right)$, $G_{1,2}\equiv 
G\left(\mathbf{x}_{1},\mathbf{x}_{2},E-\hbar\Omega\right)$.
The term in the last row, i.e. the imaginary part of the integral,
vanishes because of the GF symmetry property (\ref{eq:Geq}), and
the restriction of the integration ranges by the region inside the
solid. The obtained expression we compare with 
\begin{align*}
\mathrm{Im}I_{2} & =\mathrm{Im}\left(\iint d^{3}\mathbf{x}_{1}d^{3}
\mathbf{x}_{2}\varphi_{1}\hat{O}_{1}G_{1,2}\hat{O}_{2}\varphi_{2}^{\ast}\right)\\
 & =\iint_{\mathbf{x}_{1,2}\subset\mathcal{S}}d^{3}\mathbf{x}_{1}
 d^{3}\mathbf{x}_{2}\left[-\varphi_{1}^{\prime}\hat{O}_{1}G_{1,2}^{\prime}
 \hat{O}_{2}\varphi_{2}^{\prime\prime}+\varphi_{1}^{\prime\prime}\hat{O}_{1}
 G_{1,2}^{\prime\prime}\hat{O}_{2}\varphi_{2}^{\prime\prime}+
 \varphi_{1}^{\prime}\hat{O}_{1}G_{1,2}^{\prime\prime}\hat{O}_{2}
 \varphi_{2}^{\prime}+\varphi_{1}^{\prime\prime}\hat{O}_{1}G_{1,2}^{\prime}
 \hat{O}_{2}\varphi_{2}^{\prime}\right]\\
 & =\iint_{\mathbf{x}_{1,2}\subset\mathcal{S}}d^{3}\mathbf{x}_{1}d^{3}
 \mathbf{x}_{2}\left(\varphi_{1}^{\prime}\hat{O}_{1}G_{1,2}^{\prime\prime}
 \hat{O}_{2}\varphi_{2}^{\prime}+\varphi_{1}^{\prime\prime}\hat{O}_{1}
 G_{1,2}^{\prime\prime}\hat{O}_{2}\varphi_{2}^{\prime\prime}\right)=I,
\end{align*}
where we have again exploit the property (\ref{eq:Geq}), and the
restriction of the integration range by the volume inside the solid
due to confinement of the initial state. We thus obtain 
\begin{equation}
I=\mathrm{Im}\mathcal{G}\left(E-\hbar\Omega\right),\label{eq:I_ImG}
\end{equation}
where
\begin{align*}
\mathcal{G}\left(\hat{q},\omega\right) & \equiv\left\langle \!\left\langle 
\int d^{3}\mathbf{x}_{1}\varphi_{>}\left(\mathbf{x}_{1},\hat{q},E\right)
\hat{O}\left(\mathbf{x}_{1}\right)\hat{\psi}(\mathbf{x}_{1})|\int d^{3}
\mathbf{x}_{2}\hat{\psi}^{\dagger}(\mathbf{x}_{2})\hat{O}\left(\mathbf{x}_{2}
\right)\varphi_{>}^{\ast}\left(\mathbf{x}_{2},\hat{q},E\right)\right\rangle 
\!\right\rangle _{\omega}=\left\langle \!\left\langle \hat{C}|\hat{C}^{\dagger}
\right\rangle \!\right\rangle _{\omega}
\end{align*}
with $\hat{C}_{\sigma}$ given by Eq.\ (\ref{eq:C}) , $\omega\equiv E-\hbar\Omega$.

Now we can use the expression (\ref{eq:LEED}) for $\varphi_{>}\left(\mathbf{x}_{i},\hat{q},E\right)$
to write

\begin{align}
\hat{C} & =\sum_{\mathbf{R}_{\parallel},l,\alpha}\iint d^{2}\mathbf{x}_{\parallel}
\int dze^{-i\mathbf{k}_{\parallel}\mathbf{x}_{\parallel}}U\left(
\mathbf{x}_{\parallel},z,\hat{q},E\right)\hat{O}\left(\mathbf{x}\right)
w_{\alpha}\left[\mathbf{x}_{\parallel}-\mathbf{R}_{\parallel}
+\left(z-lc\right)\mathbf{n}-\mathbf{s}\right]a_{\mathbf{R}_{\parallel},l,\alpha}\nonumber \\
 & =\sum_{\mathbf{R}_{\parallel},l,\alpha}\iint d^{2}\mathbf{x}_{\parallel}
 \int dze^{-i\mathbf{k}_{\parallel}\left(\mathbf{x}_{\parallel}
 +\mathbf{R}_{\parallel}\right)}U\left(\mathbf{x}_{\parallel},z,\hat{q},
 E\right)\hat{O}\left(\mathbf{x}\right)w_{\alpha}\left[
 \mathbf{x}_{\parallel}+\left(z-lc\right)\mathbf{n}-\mathbf{s}\right]
 a_{\mathbf{R}_{\parallel},l,\alpha}\nonumber \\
 & =\sum_{l,\alpha}\int dz\iint d^{2}\mathbf{x}_{\parallel}\varphi_{>}
 \left(\mathbf{x},\hat{q},E\right)\hat{O}\left(\mathbf{x}\right)
 w_{\alpha}\left[\mathbf{x}_{\parallel}+\left(z-lc\right)\mathbf{n}
 -\mathbf{s}\right]\sum_{\mathbf{R}_{\parallel}}e^{-i\mathbf{k}_{\parallel}
 \mathbf{R}_{\parallel}}a_{\mathbf{R}_{\parallel},l,\alpha},\label{eq:Ckl-1}
\end{align}
and obtain Eq. (\ref{eq:Ckl}).

Substituting the Eq. (\ref{eq:Ufim}) into (\ref{eq:Ckl}) we rewrite
it in the form

\begin{align*}
\hat{C} & =\sqrt{N_{\parallel}}\sum_{m,l,\alpha}\int dz\iint d^{2}\mathbf{x}_{\parallel}e^{-i\left[\mathbf{k}_{\parallel}\mathbf{x}_{\parallel}+k_{\perp,m}^{*}\left(z-z_{0}\right)\right]}u_{m}\left(\mathbf{x},\mathbf{k}_{\parallel},E\right)\hat{O}\left(\mathbf{x}\right)w_{\alpha}\left[\mathbf{x}_{\parallel}+\left(z-lc\right)\mathbf{n}-\mathbf{s}\right]a_{\mathbf{k}_{\parallel},l,\alpha}
\end{align*}
that gives Eqs. (\ref{eq:CklM}), (\ref{eq:Ck})

For the one-band model considered in Sec. \ref{sec:surf}, and the
final sate given by Eq.(\ref{eq:LEEDmfp}) we have 
\[
\hat{C}=\sqrt{\frac{N_{\parallel}}{N_{\perp}}}\mathsf{M}\left(\mathbf{k}_{\parallel},E\right)\sum_{p}e^{ipz_{0}}\Delta_{p}a_{\mathbf{k}_{\parallel},p,\alpha}.
\]
For the GF of Eq.~Eq.(\ref{eq:Gp1p2calG}) we have 
\begin{equation}
\mathcal{G}\left(\hat{q},\omega\right)=\left|\mathsf{M}\left(\mathbf{k}_{\parallel},E\right)\right|^{2}N_{\parallel}\left\{ \frac{1}{N_{\perp}}\sum_{p}\left|\Delta_{p}\right|^{2}G_{b,\mathbf{k}}(\omega)-\frac{1}{N_{\perp}^{2}g_{\mathbf{k}_{\parallel}}(\omega)}\sum_{p_{1},p_{2}}e^{i(p_{1}-p_{2})z_{0}}\Delta_{p_{1}}\Delta_{p_{2}}^{\ast}G_{b,\mathbf{k}_{1}}(\omega)G_{b,\mathbf{k}_{2}}(\omega)\right\} ,\label{eq:G1band}
\end{equation}
where $\mathbf{k}_{i}=\mathbf{k}_{\parallel}+p_{i}\mathbf{n}$, $G_{b,\mathbf{k}}(\omega)$
is given by Eq. (\ref{eq:Gbulk}). Usual substitution $(1/N_{\perp})\sum_{p}\cdots\rightarrow(1/2\pi c)\int_{-\pi/c}^{\pi/c}\cdots dp$
gives Eq. (\ref{eq:GI1I2}). The integrals $I_{1}$ (\ref{eq:I1}),
and $I_{21}$(\ref{eq:I21}) are calculated using the substitution
\begin{equation}
z=e^{ipc},\; dz=ice^{ipc}dp,\; dp=-i\frac{dz}{cz},\label{eq:subst}
\end{equation}
then 
\begin{align}
I_{1}\left(\mathbf{k}_{\parallel},\omega\right) & =-\frac{i}{2\pi}\oint_{|z|=1}\frac{dz}{\left(1-z_{k}z\right)\left(z-z_{k}^{\ast}\right)}\frac{1}{\omega-\sigma_{\mathbf{k}_{\parallel},\omega}+T_{\mathbf{k}_{\parallel},\omega}\left(z+\frac{1}{z}\right)}\nonumber \\
 & =-\frac{i}{2\pi}\oint_{|z|=1}\frac{dz}{\left(1-z_{k}z\right)\left(z-z_{k}^{\ast}\right)}\frac{z}{T_{\mathbf{k}_{\parallel},\omega}\left(z-z_{S}\right)\left(z-z_{-S}\right)},\label{eq:I1z}
\end{align}
where $z_{k}\equiv e^{-ik_{z}^{\prime}-k_{z}^{\prime\prime}c},$ 
$\sigma_{\mathbf{k}_{\parallel},\omega}\equiv\varepsilon_{\mathbf{k}_{\parallel}}
+\Sigma_{\mathbf{k}_{\parallel},\omega}$,
\begin{equation}
z_{\pm S}=-\frac{\omega-\sigma_{\mathbf{k}_{\parallel},\omega}\mp S_{\omega}\sqrt{\left(\omega-\sigma_{\mathbf{k}_{\parallel},\omega}\right)^{2}-4T_{\mathbf{k}_{\parallel},\omega}^{2}}}{2T_{\mathbf{k}_{\parallel},\omega}},\label{eq:zs}
\end{equation}
function $S_{\omega}$ is given by Eq. (\ref{eq:Sw}).
Noting that $|z_{k}|,|z_{S}|<1$, we find two poles lying inside the
circle $|z|<1$: $z_{1}=z_{k}^{\ast}$, and $z_{2}=z_{S}$. This gives
Eq. (\ref{eq:I1R1R2}), with 
\begin{align*}
R_{1} & =\frac{1}{1-|z_{k}|^{2}}\frac{z_{k}^{\ast}}{T_{\mathbf{k}_{\parallel},\omega}\left(z_{k}^{\ast}-z_{-S}\right)\left(z_{k}^{\ast}-z_{S}\right)}\\
 & =\frac{1}{1-|z_{k}|^{2}}\frac{1}{\omega-\sigma_{\mathbf{k}_{\parallel},\omega}+T_{\mathbf{k}_{\parallel},\omega}\left(z_{k}^{\ast}+\frac{1}{z_{k}^{\ast}}\right)},\\
R_{2} & =\frac{1}{\left(1-z_{k}z_{S}\right)\left(z_{S}-z_{k}^{\ast}\right)}\frac{z_{S}}{T_{\mathbf{k}_{\parallel},\omega}\left(z_{S}-z_{-S}\right)}.
\end{align*}

Similarly, we have 
\begin{align*}
I_{21}(k_{z}^{\prime}) & =-\frac{i}{2\pi}\oint_{|z|=1}\frac{dz}{\left(1-z_{k}z\right)}\frac{z}{T_{\mathbf{k}_{\parallel},\omega}\left(z-z_{S}\right)\left(z-z_{-S}\right)}\\
 & =\frac{1}{\left(1-z_{k}z_{S}\right)}\frac{z_{S}}{T_{\mathbf{k}_{\parallel},\omega}\left(z_{S}-z_{-S}\right)}.
\end{align*}
Then 
\begin{align}
I_{2} & =\frac{1}{g_{\mathbf{k}_{\parallel}}(\omega)}I_{21}(k_{z}^{\prime})I_{21}(-k_{z}^{\prime})=\frac{1}{\left(1-z_{k}z_{S}\right)}\frac{T_{\mathbf{k}_{\parallel},\omega}\left(z_{S}-z_{-S}\right)}{\left(1-z_{k}^{\ast}z_{S}\right)}\left[\frac{z_{S}}{T_{\mathbf{k}_{\parallel},\omega}\left(z_{S}-z_{-S}\right)}\right]^{2},\nonumber \\
I & =R_{1}+R_{2}-I_{2}=\frac{1}{T_{\mathbf{k}_{\parallel},\omega}\left(1-|z_{k}|^{2}\right)\left(z_{k}^{*}z_{S}-1\right)\left(z_{-S}-z_{k}\right)}\label{eq:R2mI2}
\end{align}
Above, we have taken into account that 
\begin{align}
g_{\mathbf{k}_{\parallel}}(\omega) & =\frac{1}{T_{\mathbf{k}_{\parallel},\omega}\left(z_{S}-z_{-S}\right)}=\frac{S_{\omega}}{\sqrt{\left(\omega-\sigma_{\mathbf{k}_{\parallel},\omega}\right)^{2}-4T_{\mathbf{k}_{\parallel},\omega}^{2}}}.\label{eq:gkap}
\end{align}
Substituting $z_{k}$ and $z_{\pm S}$ into the above expressions,
we obtain the formulas for various contributions into (\ref{eq:GI1I2}):
\begin{align}
R_{1}\left(\mathbf{k}_{\parallel},\omega\right) & =\left\{ \left(1-e^{-2k_{z}^{\prime\prime}c}\right)\left[\omega-\sigma_{\mathbf{k}_{\parallel},\omega}+2T_{\mathbf{k}_{\parallel},\omega}\cos\left(k_{z}^{\prime}c+ik_{z}^{\prime\prime}c\right)\right]\right\} ^{-1},\label{eq:R1}\\
R_{2}\left(\mathbf{k}_{\parallel},\omega\right) & =\frac{S_{\omega}T_{\mathbf{k}_{\parallel},\omega}e^{k_{z}^{\prime\prime}c}}{\left[2T_{\mathbf{k}_{\parallel},\omega}\cosh k_{z}^{\prime\prime}c+\left(\omega-\sigma_{\mathbf{k}_{\parallel},\omega}\right)\cos k_{z}^{\prime}c+iS_{\omega}\sin k_{z}^{\prime}c\sqrt{\left(\omega-\sigma_{\mathbf{k}_{\parallel},\omega}\right)^{2}-4T_{\mathbf{k}_{\parallel},\omega}^{2}}\right]\sqrt{\left(\omega-\sigma_{\mathbf{k}_{\parallel},\omega}\right)^{2}-4T_{\mathbf{k}_{\parallel},\omega}^{2}}},\label{eq:R2}\\
I_{2}\left(\mathbf{k}_{\parallel},\omega\right) & =\frac{2S_{\omega}T_{\mathbf{k}_{\parallel},\omega}^{2}}{\left[\left(\omega-\sigma_{\mathbf{k}_{\parallel},\omega}+S_{\omega}\sqrt{\left(\omega-\sigma_{\mathbf{k}_{\parallel},\omega}\right)^{2}-4T_{\mathbf{k}_{\parallel},\omega}^{2}}\right)\left(\omega-\sigma_{\mathbf{k}_{\parallel},\omega}+2T_{\mathbf{k}_{\parallel},\omega}e^{-k_{z}^{\prime\prime}c}\cos k_{z}^{\prime}c\right)-2T_{\mathbf{k}_{\parallel},\omega}^{2}\left(1-e^{-2k_{z}^{\prime\prime}c}\right)\right]}\label{eq:I2kap}\\
 & \times\frac{1}{\sqrt{\left(\omega-\sigma_{\mathbf{k}_{\parallel},\omega}\right)^{2}-4T_{\mathbf{k}_{\parallel},\omega}^{2}}}.\nonumber 
\end{align}
On the other hand, we note that 
\begin{align*}
z_{S} & =-T_{\mathbf{k}_{\parallel},\omega}g_{s}\left(\omega-\sigma_{\mathbf{k}_{\parallel},\omega},T_{\mathbf{k}_{\parallel},\omega}\right),\quad z_{-S}=-\frac{\omega-\sigma_{\mathbf{k}_{\parallel},\omega}}{T_{\mathbf{k}_{\parallel},\omega}}-z_{S},
\end{align*}
then the denominator of the second fraction of the right-hand side
of (\ref{eq:R2mI2}) is 
\begin{align*}
 & z_{k}^{*}z_{S}z_{-S}-|z_{k}|^{2}z_{S}-z_{-S}+z_{k}=\\
 & =\frac{\omega-\sigma_{\mathbf{k}_{\parallel},\omega}+2T_{\mathbf{k}_{\parallel},\omega}e^{-k_{z}^{\prime\prime}c}\cos k_{z}^{\prime}c-T_{\mathbf{k}_{\parallel},\omega}^{2}\left(1-e^{-2k_{z}^{\prime\prime}c}\right)g_{s}\left(\omega-\sigma_{\mathbf{k}_{\parallel},\omega},T_{\mathbf{k}_{\parallel},\omega}\right)}{T_{\mathbf{k}_{\parallel},\omega}},
\end{align*}
and $\mathcal{G}$ is given by Eq. (\ref{eq:Gfin}).

The behavior near the resonance frequency $\omega_{r}$ is described
by the expression 
\begin{align}
F_{r}\left(\mathbf{k}_{\parallel},\omega\right) & =\frac{Z\left(\omega_{r}\right)e^{k_{z}^{\prime\prime}c}}{\left(\omega-\omega_{r}\right)\cosh k_{z}^{\prime\prime}c+i\Gamma}\label{eq:Frfull}\\
\Gamma & =2Z\left(\omega_{r}\right)\left|T_{\mathbf{k}_{\parallel},\omega_{r}}\right|\tanh k_{z}^{\prime\prime}c\sqrt{\sinh^{2}k_{z}^{\prime\prime}c+\sin^{2}k_{z}^{\prime}c}\label{eq:Gamfull}
\end{align}

\end{widetext}

\bibliography{mh}

\end{document}